\documentclass[preprint,preprintnumbers,amsmath,amssymb,floatfix,nofootinbib]{revtex4}
\usepackage{epsfig}
\usepackage{amsmath}
\usepackage{amssymb}
\usepackage{amsfonts}
\usepackage{multirow}
\usepackage{rotating}
\usepackage{verbatim}
\usepackage{subfigure}
\usepackage{float}
\def\br(#1,#2){\left\langle#1#2\right\rangle}
\def\sq(#1,#2){\left[#1#2\right]}
\def\s(#1,#2){s_{#1 #2}}
\def\t(#1,#2,#3){s_{#1 #2 #3}}

\begin{document}



\title{Combining NLO QCD and Electroweak Radiative Corrections to
  $W$-boson Production at Hadron Colliders in the {\tt POWHEG} Framework} 
\author{C.~Bernaciak} 
\email{C.Bernaciak@ThPhys.Uni-Heidelberg.de}
\affiliation{Institut f\"ur Theoretische Physik, Universit\"at Heidelberg, D-69120 Heidelberg, Germany}
\author{D.~Wackeroth}
\email{dow@ubpheno.physics.buffalo.edu}
\affiliation{Department of Physics, SUNY at Buffalo, Buffalo, NY 14260-1500, U.S.A.}
\date{\today}
\begin{abstract}
  The precision measurement of the mass of the $W$ boson is an important goal
  of the Fermilab Tevatron and the CERN Large Hadron Collider (LHC). It
  requires accurate theoretical calculations which incorporate both
  higher-order QCD and electroweak corrections, and also provide an interface
  to parton-shower Monte Carlo programs which make it possible to
  realistically simulate experimental data. In this paper, we present a
  combination of the full ${\cal O}(\alpha)$ electroweak corrections of {\tt
    WGRAD2}, and the next-to-leading order QCD radiative corrections to
  $W\to\ell\nu$ production in hadronic collisions in a single event generator
  based on the {\tt POWHEG} framework, which is able to interface with the
  parton-shower Monte Carlo programs {\tt Pythia} and {\tt Herwig}. Using this
  new combined QCD+EW Monte Carlo program for $W$ production, we provide numerical results for total cross sections and
  kinematic distributions of relevance to the $W$ mass measurement at the
  Tevatron and the LHC for the processes $pp,p\bar p \to W^\pm \to \mu^\pm
  \nu_\mu$. In particular, we discuss the impact of EW corrections in the
  presence of QCD effects when including detector resolution effects.
   
\end{abstract}
%
\maketitle

\section{Introduction}\label{sec:intro}

The precision measurement of the mass of the $W$-boson, $M_W$, is an
important goal of the Fermilab
Tevatron~\cite{Aaltonen:2007ps,Abazov:2009cp,:2009nu} and the CERN Large
Hadron Collider (LHC). With precise knowledge of $M_W$ and the top
quark mass, $m_t$, indirect information on the mass of the Higgs
boson, $M_H$, within the Standard Model (SM) can be extracted from the
$M_H$ dependence of radiative corrections to the $W$
mass. Global SM fits to all electroweak precision data performed by
the {\tt GFITTER} collaboration and the LEP Electroweak Working Group
predict the SM Higgs mass to be $M_H=96^{+31}_{-24}$~GeV ($1
\sigma$)~\cite{Flacher:2008zq,Baak:2011ze} and
$M_H=92^{+34}_{-26}$~GeV ($68\%$~C.L.)~\cite{Alcaraz:2009jr},
respectively, which lies well within the mass range presently probed
by the Tevatron~\cite{CDFandDO:2011aa} and
LHC~\cite{Chatrchyan:2011tz,Aad:2011qi} experiments. Future more
precise measurements of the $W$ and top quark masses together with
improvements in the SM predictions of $M_W$ are expected to
considerably improve the indirect determination of $M_H$. At the
Tevatron, for an integrated luminosity of ${\cal L}=10~\rm{fb}^{-1}$,
an ultimate precision of $\delta M_W=$15~MeV for the $W$ mass may be
possible~\cite{Kotwal:2008zz}. For the LHC, estimates range from
$\delta M_W=7$~MeV~\cite{Besson:2008zs} to $\delta
M_W=20$~MeV~\cite{Buge:2006dv} for ${\cal L}=10~{\rm fb}^{-1}$,
depending on the assumptions made for detector resolutions and
theoretical uncertainties.  With a dedicated
program~\cite{Krasny:2010vd}, one may be able to achieve $\delta
M_W={\cal O}(10$~MeV).

In hadronic collisions, the $W$ boson mass can be determined from the
transverse mass distribution of the lepton pair, $M_T(l\nu)$,
originating from the $W$ decay, $W\to\ell\nu$, and the transverse
momentum distribution of the charged lepton or neutrino.  Both QCD and
electroweak (EW) corrections play an important role in the measurement
of $W$ observables at hadron colliders. It is imperative to
control predictions for observables relevant to $W$ production at least at the $1\%$ level.
Also the transverse momentum distribution of the $W$ boson is
an important ingredient in the current $W$ mass measurement at the
Tevatron (see, e.g. Ref.~\cite{Kotwal:2008zz} for a review). 
In lowest order (LO) in perturbation theory, the $W$ boson
is produced without any transverse momentum. Only when QCD corrections
are taken into account does the $W$ boson acquire a non-negligible
transverse momentum, $p_T^W$. For a detailed understanding of the
$p_T^W$ distribution, it is necessary to resum the soft gluon emission
terms, and to model non-perturbative QCD corrections. This has been done either
by using calculations targeted specifically for resummation and
parametrizing non-perturbative effects (see
e.g. Refs.~\cite{Balazs:1997xd} and~\cite{Landry:1999an}), or
interfacing a calculation of $W$ boson production at next-to-leading
order (NLO) in QCD with a parton-shower Monte Carlo (MC) program and
tuning the parameters used to describe the non-perturbative
effects. This approach has been pursued in
Refs.~\cite{Frixione:2002ik,Alioli:2008gx,Hamilton:2008pd}, for instance.  Fixed
higher-order predictions beyond NLO are known for fully differential distributions 
through next-to-next-to-leading 
order in QCD~\cite{Anastasiou:2003ds,Melnikov:2006di,Catani:2009sm}, and
recently first steps towards a calculation of the complete mixed EW-QCD ${\cal O}(\alpha_s
\alpha)$ corrections to the Drell-Yan production process were made in Ref.~\cite{Kilgore:2011pa}.

While QCD corrections only indirectly affect the $W$ mass extracted
from the $M_T(l\nu)$ distribution, EW radiative corrections can
considerably distort the shape of this distribution in the region
sensitive to the $W$ mass. For instance, final-state photon radiation
is known to shift $M_W$ by ${\cal O}(100$~MeV)~
\cite{Aaltonen:2007ps,Abazov:2009cp,cdfwmass,d0wmass,unknown:2003sv,
Abe:1994qn,Affolder:2000mt,Abazov:2002xj}.  In the last few years,
significant progress in our understanding of the EW corrections to $W$
boson production in hadronic collisions has been made. The complete
${\cal O}(\alpha)$ EW radiative corrections to
$p\,p\hskip-7pt\hbox{$^{^{(\!-\!)}}$} \to W^{\pm} \to\ell^{\pm} \nu$
($\ell=e,\,\mu$) were calculated by several
groups~\cite{Wackeroth:1996hz,Baur:1998kt,Dittmaier:2001ay,Baur:2004ig,
Arbuzov:2005dd,CarloniCalame:2006zq,Zykunov:2008zz} and found to agree
~\cite{Buttar:2006zd,Gerber:2007xk}. First steps towards going beyond
fixed-order in QED radiative corrections in $W$ production were taken
in
Refs.~\cite{CarloniCalame:2003ux,Placzek:2003zg,Golonka:2005pn,Hamilton:2006xz,Brensing:2007qm}, for instance,
by including the effects of final-state multiple photon radiation. For
a review of the state-of-the-art of predictions for $W$ production at
hadron colliders see, e.~g.,
Refs.~\cite{Buttar:2006zd,Gerber:2007xk,Laenen:2009zz}.

As a result of all these studies, given the anticipated accuracy of a $W$
boson mass measurement at the Tevatron and the LHC, it has become increasingly
clear that it is necessary to not only fully understand and control the
separate higher-order QCD and EW corrections, but also their combined effects.
A first study of combined effects can be found in Ref.~\cite{Cao:2004yy},
where final-state photon radiation was added to a calculation of $W$ boson
production which includes NLO and resummed QCD corrections. This study showed
that the difference in the effects of EW corrections in the presence of QCD
corrections and of simply adding the two predictions may be not negligible in
view of the anticipated precision. Moreover, in the relevant kinematic region,
i.e. around the Jacobian peak, the QCD corrections tend to compensate some of
the effects of the EW corrections.  In Ref.~\cite{Balossini:2009sa} the full
set of EW ${\cal O}(\alpha)$ corrections of {\tt
  HORACE}~\cite{CarloniCalame:2006zq} and the QCD NLO corrections to $W$
production were combined in the {\tt MC@NLO} framework~\cite{Frixione:2002ik}
which is interfaced with the parton-shower MC program {\tt
  Herwig}~\cite{Corcella:2000bw}. The results of a combination of the EW
${\cal O}(\alpha)$ corrections to $W$ production as implemented in
{\tt SANC}~\cite{Arbuzov:2005dd} with {\tt
  Pythia}~\cite{Sjostrand:2006za} and {\tt Herwig} can be found in
Ref.~\cite{Richardson:2010gz}, without, however, performing a matching of NLO
QCD corrections to the parton shower.

In this paper, we present a combination of the full EW ${\cal O}(\alpha)$
radiative corrections of Ref.~\cite{Baur:1998kt,Baur:2004ig} contained in the
public MC code {\tt WGRAD2} and the QCD corrections to $W\to\ell\nu$
production of {\tt POWHEG-W}~\cite{Alioli:2008gx}.  One advantage of the {\tt
  POWHEG} method~\cite{Nason:2004rx,Frixione:2007vw,Alioli:2010xd} for the use
in a detector simulation is that it only generates positive weighted events.
Moreover, it provides an interface to both {\tt Herwig} and {\tt Pythia}.  It
is well suited as a starting point for combining EW and QCD corrections to
$W$-boson production in one MC program to serve as an analysis tool in the
$W$-mass measurement of the Tevatron and LHC experiments. The resulting MC
code, called in the following {\tt POWHEG-W\_EW}, is publicly available at the {\tt POWHEG Box}
webpage~\cite{webpage} and allows the simultaneous study of the effects of
both QCD and NLO EW corrections with both {\tt Pythia} and {\tt Herwig}.
We do not include the effects of photon-induced processes and of multiple photon
radiation. As has been found in earlier
studies~\cite{CarloniCalame:2003ux,Placzek:2003zg,Golonka:2005pn,Hamilton:2006xz,Brensing:2007qm,Gerber:2007xk,Richardson:2010gz},
both effects, although small, still can have a non-negligible impact on the
$W$-mass measurement and should be included in view of the anticipated final
precision of the $M_W$ measurement at the Tevatron. This is left to a future
publication.

The technical details of our calculation and implementation of EW ${\cal
  O}(\alpha)$ corrections in {\tt POWHEG-W} are described in
Section~\ref{sec:theory}. In Section~\ref{sec:results} we first describe our
crosschecks, and then present numerical results for total cross sections and
distributions which are of interest for the $W$-mass measurement at the
Tevatron and the LHC.  In particular, we study the combined effects of EW
${\cal O}(\alpha)$ and QCD corrections on the $M_T(\mu \nu_\mu)$ and
$p_T(\mu)$ distributions in $pp,p\bar p \to W^\pm \to \mu^\pm \nu_\mu$, taking
into account detector resolution effects and using {\tt Pythia} to simulate
parton showering.  Finally, our conclusions are presented in
Section~\ref{sec:conclusions}.

\section{Theoretical Framework}\label{sec:theory}
In the following we concentrate on describing our implementation of the
complete EW ${\cal O}(\alpha)$ corrections to $W$ production via the Drell-Yan
mechanism $q_i \overline{q}_{i'}\to W \rightarrow f \bar f' (\gamma)$ in {\tt
  POWHEG-W}. We refer to the literature for a detailed description of QCD and
EW corrections to $W$ boson production at hadron colliders as implemented in
{\tt POWHEG-W}~\cite{Alioli:2008gx} and {\tt
  WGRAD2}~\cite{Baur:1998kt,Baur:2004ig}, respectively. To illustrate our
implementation, we start with a schematic presentation of the parton-level NLO
QCD cross section to $W$ production as given in Ref.~\cite{Nason:2004rx} (see
also Refs.~\cite{Frixione:2007vw,Alioli:2010xd} for a detailed description of
the {\tt POWHEG BOX}):
\begin{equation}
d\sigma=B(\Phi_2)d\Phi_2 + V(\Phi_2)d\Phi_2 + [R(\Phi_3)d\Phi_3-C(\Phi_3)d\Phi_3 P]
\end{equation}
where the $2\to 3$ phase space of the radiated parton is given by
$d\Phi_3=d\bar \Phi_2 d\Phi_{\text{rad}}$, $B, V, R$ denote the Born, virtual
and real emission contributions, respectively, and $C$ are counter terms,
derived in a suitable subtraction scheme, that ensure that the term in the
square brackets is non-singular.  $P$ denotes a projection of $2\to 3$
kinematics onto $2\to 2$ kinematics. After some manipulation suitable for
interfacing $d\sigma$ with a parton-shower MC, the cross section can be
written as follows~\cite{Nason:2004rx}:
\begin{equation}\label{eq:nason}
  d\sigma=\bar B(\Phi_2)d\Phi_2 [ \Delta_R^{NLO}(0)+ \Delta_R^{NLO}(p_T) \frac{R(\Phi_2,\Phi_{\text{rad}})}{B(\Phi_2)} d\Phi_{\text{rad}}]
\end{equation}
where the term in square brackets contains the Sudakov form factor
$\Delta_R^{NLO}$ and generates the first emission of a light parton, while all
subsequent emissions are handled by the parton-shower MC. $\bar B$ is defined
as~\cite{Nason:2004rx}
\begin{equation}
\bar B(\Phi_2)=
B(\Phi_2)+ V(\Phi_2) + \int [R(\Phi_2,\Phi_{\text{rad}})-C(\Phi_2,\Phi_{\text{rad}})] d\Phi_{\text{rad}}
\end{equation}  
and offers a straightforward way of adding the EW corrections to the QCD
${\cal O}(\alpha_s)$ corrections contained in $V$ and $R$.  In detail, all
changes made to {\tt POWHEG-W} in order to include the ${\cal O}(\alpha)$ EW
corrections of {\tt WGRAD2} are portrayed by the boxed terms contained in the
$\bar{B}^{f_b}(\Phi_2)$ portion of {\tt POWHEG-W} as follows (now we follow
the notation of Ref.~\cite{Frixione:2007vw}):
\begin{eqnarray} \label{eq:main}
\bar{B}^{f_b}(\Phi_2)&=&\left[B(\Phi_2)+V_{\text{QCD}}(\Phi_2)+\boxed{V_{\text{EW}}(\Phi_2)}
\,\right]_{f_b} \\ \nonumber
&+&\sum_{\alpha_r=0}^2\int\left\{d\Phi_{\text{rad}}\left[R_{\text{QCD}}
(\Phi_{3})-C(\Phi_{3})\right]\right\}_{\alpha_r,f_b}^{\bar \Phi_2^{\alpha_r}=\Phi_2} \\ \nonumber
&+& 
\boxed{\int d\Phi^{\alpha_r\scriptscriptstyle{=0}}_{\text{rad}}R_{\text{EW}}^{f_b}(\Phi_{3}) 
\theta(E_\gamma-\delta_s \frac{\sqrt{\hat s}}{2}) \theta(\hat s_{q\gamma}-\delta_c E_\gamma \sqrt{\hat s}) \theta(\hat s_{\bar q\gamma}-\delta_c E_\gamma \sqrt{\hat s})} \\ \nonumber
&+&\int\frac{dz}{z}\left[\sum_{\alpha_{\oplus}=1}^2G_{\text{QCD},\oplus}^{\alpha}(\Phi_{2,\oplus})
+\,\boxed{G^1_{\text{EW},\oplus}(\Phi_{2,\oplus}) \theta(1-\delta_s-z)}\,\right]_{f_b} \\ \nonumber
&+&\int\frac{dz}{z}\left[\sum_{\alpha_{\ominus}=1}^2G_{\text{QCD},\ominus}^{\alpha}(\Phi_{2,\ominus})
+\,\boxed{G^1_{\text{EW},\ominus}(\Phi_{2,\ominus}) \theta(1-\delta_s-z)}\,\right]_{f_b}
\end{eqnarray}
where the subscript `QCD' refers to the original \texttt{POWHEG-W} terms.  For
the real terms, $\alpha_r =0$ corresponds to singularities occurring when the
initial-state emitters are $q$ or $\bar{q}^{\scriptscriptstyle{\prime}}$ and
the gluon could be emitted from either of them.  $\alpha_r =1$ corresponds to
a gluon emitting an antiquark, $\bar{q}$ or
$\bar{q}^{\scriptscriptstyle{\prime}}$, and $\alpha_r =2$ to a gluon emitting
a quark $q^{\scriptscriptstyle{\prime}}$ or $q$. The $f_b$ correspond to each
particular flavor structure at the Born level where in the case of
$W\to\ell\nu$ production there are twelve.

For each collinear piece, $\alpha_{\oplus(\ominus)}=1$ corresponds to a
quark/antiquark from a hadron with positive (negative) rapidity emitting a
collinear gluon and $\alpha_{\oplus(\ominus)}=2$ to a positive(negative)
rapidity gluon emitting a collinear quark/antiquark.

In order to incorporate real photon emission as part of the EW ${\cal
  O}(\alpha)$ corrections, the same momentum used to denote the radiated
parton (gluon or quark/antiquark) is used to denote the radiated photon.
However, because our implementation of the ${\cal O}(\alpha)$ EW corrections
does not include photon-induced processes, the EW contribution to the real
term of Eq.~\ref{eq:main} is incorporated only into the $\alpha_r=0$
contribution and likewise for the collinear terms there is only an
$\alpha_{\oplus,\ominus}=1$ term, as denoted in Eq.~\ref{eq:main}.

As described in detail in Ref.~\cite{Baur:1998kt} (see also
Ref.~\cite{Harris:2001sx}) we use the phase space slicing (PSS) method to
extract the soft and collinear singular regions in the contribution of real
photon radiation described by $R_{\text{EW}}^{f_b}$. In these regions the
integration over the photon phase space is performed analytically using a soft
and collinear approximation of $R_{\text{EW}}^{f_b}$, which is valid as long
as the PSS parameters $\delta_s$ and $\delta_c$ are chosen to be sufficiently
small. The soft part is included in $V_{EW}$ and the remnant of the
initial-state collinear singularity after mass factorization is denoted by
$G_{EW,(\ominus,\oplus)}^1$. Explicit expressions for these contributions and
a detailed description of the QED factorization scheme can be found in
Ref.\cite{Baur:1998kt}.  Finally, we refer to the appendix for the details of
this implementation of EW ${\cal O}(\alpha)$ corrections into {\tt POWHEG-W}.

\section{Numerical Results}\label{sec:results}
Numerical evaluations of the total cross sections and relevant $W$ boson
distributions have been obtained for both Tevatron ($\sqrt{s}=1.96$ TeV) and
LHC ($\sqrt{s}=7$ TeV) processes using both {\tt Pythia} and {\tt Herwig} for
QCD parton showering.  After a description of the numerical setup in
Section~\ref{sec:setup}, and of crosschecks performed to validate the
implementation of the EW corrections of {\tt WGRAD2} in {\tt POWHEG-W} in
Section~\ref{sec:checks}, we provide results for the total inclusive cross
sections with and without parton showering and/or EW corrections in
Section~\ref{sec:totcs}.  $W$ boson observables that are relevant to EW
precision studies, specifically to the measurement of $M_W$, are defined and
their distributions shown in Section~\ref{sec:distr}.
\subsection{Setup}\label{sec:setup}
The setup used to obtain the results presented in this paper 
closely follows Ref.~\cite{Gerber:2007xk}:
\begin{itemize}
\item[$\bullet$]{\textbf{SM input parameters:}}
	\begin{itemize}
	\item[$\bullet$]{\textbf{masses:} $M_Z=91.1876$ GeV, $M_W=80.398$ GeV, 
$M_H=115$ GeV\\
                         $m_e=0.51099891$ MeV, $m_\mu=0.1056583668$ GeV, $m_\tau=1.77684$
                         GeV\\
$m_u=0.06983$~keV, $m_c=1.2$~GeV, $m_t=171.2$~GeV \\
$m_d=0.06984$~keV, $m_s=0.15$~GeV, $m_b=4.6$~GeV }
	\item[$\bullet$]{\textbf{$W$ width:} $\Gamma_W=2.141$ GeV}
	\item[$\bullet$]{\textbf{EW coupling parameters:} 
$\alpha(0)=1/137.035999679$\\
	$\cos\theta_w=M_W/M_Z$, $\sin^2\theta_w=1-\cos^2\theta_w$}
\item[$\bullet$]{\textbf{CKM matrix elements:} $|V_{ud}|=|V_{cs}|=0.975$, 
                        $|V_{us}|=|V_{cd}|=0.222$,\\
                        $|V_{ub}|=|V_{cb}|=|V_{td}|=|V_{ts}|=0, |V_{tb}|=1$}
	\end{itemize}
\item[$\bullet$]{\textbf{WGRAD2 flags:} qnonr=0, QED=4, lfc=1}
\item[$\bullet$]{\textbf{renormalization/factorization scales:} $\mu_F=\mu_R=\mu=W$ boson invariant mass ($\mu=\mu_{QED}=\mu_{QCD}$)}
\item[$\bullet$]{\textbf{Parton Distribution Function (PDF):} CTEQ10~\cite{cteq:10}}
\item[$\bullet$]{\textbf{bare acceptance cuts:} $p_{T}(l) > 25$ GeV, $p_{T}(\nu_l) > 25$ 
                                                    GeV, and $|\eta_{\ell}| < 1$ }
\item[$\bullet$]{\textbf{calometric setup:}
in addition to bare acceptance cuts, smearing of the four-momenta is applied, and
we limit the photon energy for small muon-photon angles as described below. }
\item[$\bullet$]{\textbf{{\tt Pythia} settings:} MSTP(61)=1, MSTP(71)=1, MSTJ(41)=1 which
                                    corresponds to all QED showering turned off} 
\end{itemize}
The calometric setup includes smearing of the final-state four-momenta to take
into account the uncertainty in the energy measurement in the detector.
Gaussian smearing of the final-state four-momenta is simulated with D0 or
ATLAS inspired smearing routines.  All observables are then calculated from
the smeared momenta. Muons are detected in the muon chamber and the
requirement that the associated track is consistent with a minimum ionizing
particle. Therefore, for muons at the Tevatron and the LHC, we require a small
photon energy for small muon-photon opening angles, i.e. we require that
$E_\gamma<2$ GeV for $\Delta R_{\mu\gamma} < 0.1$ and $E_\gamma<0.1E_\mu$ 
for $0.1<\Delta R_{\mu\gamma} < 0.4$.  $\Delta R_{l\gamma}$ denotes the
separation of a charged lepton and photon in the pseudo-rapidity azimuthal
angle plane defined as:
\begin{equation}
\Delta R_{l\gamma} = \sqrt{\Delta\phi^2_{l\gamma} + \Delta\eta^2_{l\gamma}}
\end{equation}

The results in this paper are obtained in the constant-width scheme and by
using the fine structure constant, $\alpha(0)$, in both the LO and NLO EW
calculation of the $W$ observables.  Since QED radiation has the dominant
effect on observables relevant to the $W$ mass measurement, we only include
resonant weak corrections (qnonr=0), i.e. we neglect weak box diagrams. Their
impact is important in kinematic distributions away from the resonance region
and can be studied by choosing qnonr=1. We include the full set of QED
contributions (QED=4), i.e. initial-state and final-state radiation as well as
interference contributions. The QED and QCD factorization and QCD
renormalization scales are chosen to be equal and we assume that the
factorization of the photonic initial-state quark mass singularities is done
in the QED DIS scheme (lfc=1). The QED $\overline{\rm MS}$ scheme is
implemented as well (lfc=0) and both schemes are defined in analogy to the
corresponding QCD factorization schemes. A description of the QED
factorization scheme as implemented in {\tt POWHEG-W\_EW} can be found in
Ref.~\cite{Baur:1998kt}.
  
The fermion masses only contribute to the EW gauge boson self-energies and as
regulators of the collinear singularity.  The mass of the charged lepton is
included in the phase space generation of the final-state four-momenta and
serves as a regulator of the singularity arising from collinear photon
radiation off the charged lepton. Thus, no collinear cut needs to be applied
(collcut=0 in {\tt POWHEG-W\_EW}) on final-state photon radiation, allowing
the study of finite lepton-mass effects.  Note that the application of a
collinear cut on final-state photon radiation (collcut=1) is only allowed in
the electron case and only when a recombination of the electron and photon
momenta is performed in the collinear region (usually defined by $\Delta
R_{e\gamma}<R_{cut}$, see Ref.~\cite{Baur:1998kt} for a detailed discussion).
In this paper we present results for the $pp,p\bar p \to W^\pm \to \mu^\pm
\nu_\mu$ processes in both the bare and calometric setup.
    
\subsection{Crosschecks}\label{sec:checks}
In order to be sure that the EW corrections are properly implemented, a number
of crosschecks were performed.  In the first, the QCD corrections in {\tt
  POWHEG-W\_EW} were turned off (i.e. all terms labeled with the subscript
`QCD' in Eq.~\ref{eq:main} were set to zero) and the numerical results of each
piece of the NLO EW corrections (i.e. $V_{\text{EW}},
G_{\text{EW},(\ominus,\oplus)}^1, R_{\text{EW}}$ of Eq.~\ref{eq:main}) were
compared to {\tt WGRAD2}. We also compared results for the total inclusive
cross section and the $M_T(W)$ and $p_T(\ell)$ distributions obtained with
{\tt POWHEG-W\_EW} when only including EW ${\cal O}(\alpha)$ corrections with
those obtained with {\tt WGRAD2}.  In all these comparisons we found good
agreement within the statistical uncertainties of the numerical integration
(see also Sections~\ref{sec:totcs},\ref{sec:distr}).  This is the primary
indication that the NLO EW corrections were implemented properly using the
numerical phase space integration of the \texttt{POWHEG BOX}.  In the second
type of crosscheck, the numerical cancellation of the PSS parameters
$\delta_s$ and $\delta_c$ was tested by running {\tt POWHEG-W\_EW} without the
NLO QCD corrections (or parton showering capabilities) of {\tt POWHEG-W} for
different choices of these parameters and observing that the cross sections
agree within the statistical error of the numerical integration as long as the
PSS parameters are chosen small enough so that the soft/collinear
approximation is valid. To illustrate this cancellation we show in
Table~\ref{tab:psscx} the results for the total inclusive cross sections for
$W^\pm \to \mu^\pm \nu_\mu$ production at the Tevatron and the LHC.  Finally,
we also checked that the QCD NLO cross sections still coincide with those
obtained with the original code, {\tt POWHEG-W}.

\begin{table}[H]
\begin{center}
\begin{tabular}{c|c|c|c|}
\cline{2-4}
 & Tevatron & \multicolumn{2}{|c|}{LHC} \\
\hline
\multicolumn{1}{|c|}{($\delta_s$,$\delta_c$)} & $W^+$ & $W^+$ & $W^-$ \\
\hline
\multicolumn{1}{|c|} {0.01,0.005}  & 362.4(2) & 1059.0(5) & 758.7(8)\\ 
\hline
\multicolumn{1}{|c|}{0.01,0.001}  & 362.4(2) &  1059.1(7) & 759.2(5)\\ 
\hline
\multicolumn{1}{|c|}{0.001,0.0005}  & 362.3(2) &  1059.4(9) & 759.4(5)\\ 
\hline
\multicolumn{1}{|c|}{0.001,0.0001}  & 362.3(2) &  1059.2(8) & 759.3(5)\\ 
\hline
\end{tabular}
\caption{Results for the total cross sections (in pb) to $pp,p\bar p\to W^\pm
  \to\mu^\pm \nu_\mu$ for different choices of $\delta_s$ and $\delta_c$
  parameters at the Tevatron ($\sqrt{S}=1.96$ TeV) and LHC ($\sqrt{S}=7$ TeV).
  These results reflect the exact (weighted) NLO results of {\tt POWHEG-W\_EW}
  when only including EW corrections, with bare cuts.}
\label{tab:psscx}
\end{center}
\end{table}

\subsection{Total Cross Sections}\label{sec:totcs}
In Tables~\ref{tab:totcsbare} and \ref{tab:totcscalo} we present results
obtained with {\tt POWHEG-W\_EW} for the total cross sections of $pp,p\bar
p\to W^\pm \to \mu^\pm \nu_\mu$ processes at the Tevatron with center-of-mass
(CM) energy $\sqrt{S}=1.96$~TeV and the LHC with $\sqrt{S}=7$~TeV for the bare
and calometric setup, respectively. We show results separately for the NLO EW
and NLO QCD cross sections and combined NLO EW+QCD results with and without
QCD parton shower for both {\tt Pythia} and {\tt Herwig}.  The NLO EW and QCD
results coincide with those that can be obtained with {\tt WGRAD2} and {\tt
  POWHEG-W}, respectively.  At the level of the total cross sections, the
combined results can be approximated by simply adding the QCD and EW cross
sections.  As we will see in Section~\ref{sec:distr}, this is not necessarily
the case when studying kinematic distributions after applying parton
showering.  For instance, for $W^+$ production at the Tevatron(LHC) with bare
cuts (see Table ~\ref{tab:totcsbare}) the EW ${\cal O}(\alpha)$ corrections
increase the LO total cross section by 3.6\%(3.4\%), the combined (QCD+EW)
corrections increase the QCD cross section at NLO by 3.3\%(3.5\%) and when
parton-showering with {\tt Pythia} is included by $3.7\pm 0.4\%(3.7\pm 0.4\%)$.  When
considering the calometric setup (see Table ~\ref{tab:totcscalo}) the impact
of the NLO EW corrections is considerably reduced, and the EW ${\cal
  O}(\alpha)$ corrections increase the LO total cross section by only
1.6\%(1.4\%), the combined (QCD+EW) corrections increase the QCD cross section
at NLO by 1.1\%(1.1\%) and when parton-showering with {\tt Pythia} is included by $1.6\pm 0.3\%
(0.9 \pm 0.3\%)$.

\begin{table}[H]
\begin{center}
\begin{tabular}{c|c|c|c|c|c|c|}
\cline{2-7} 
& \multicolumn{2}{|c|}{Tevatron} & \multicolumn{4}{|c|}{LHC}\\
\cline{2-7}
& \multicolumn{2}{|c|}{$W^+$} & \multicolumn{2}{|c|}{$W^+$} &\multicolumn{2}{|c|}{$W^-$}\\
\hline
\multicolumn{1}{|c|}{LO} & \multicolumn{2}{|c}{ 349.81(2)[349.77(1)] } &
\multicolumn{2}{|c|}{1024.0(1)[1023.9(1)] }& \multicolumn{2}{c|}{731.69(6)[731.63(2)] } \\   
\hline
\multicolumn{1}{|c|}{NLO EW} & \multicolumn{2}{|c|}{ 362.4(2)[362.55(2)]} & \multicolumn{2}{|c|}{1059.0(5)[1059.6(1)] }&
\multicolumn{2}{c|}{758.7(8)[759.26(3)]}\\   
\hline
\multicolumn{1}{|c|}{NLO QCD} & \multicolumn{2}{|c|}{384.66(4) } & \multicolumn{2}{|c|}{1022.7(2) }& \multicolumn{2}{|c|}{750.8(1) } \\   
\hline
\multicolumn{1}{|c|}{NLO (QCD+EW)} & \multicolumn{2}{|c|}{397.2(2) } & \multicolumn{2}{|c|}{1058.0(6) }& \multicolumn{2}{|c|}{778(1)} \\   
\hline
& \texttt{Pythia} & \texttt{Herwig}  & \texttt{Pythia} & 
\texttt{Herwig} & \texttt{Pythia} & \texttt{Herwig} \\
\hline
\multicolumn{1}{|c|}{LO$\otimes$PS}  & 308.4(7) & 311.8(7) & 854(3) & 866(3) & 634(2) & 639(2) \\
\hline
\multicolumn{1}{|c|}{NLO QCD$\otimes$PS} & 375.3(8) & 378.5(8) & 1014(3) & 1027(3) & 744(2) & 750(2) \\
\hline
\multicolumn{1}{|c|}{NLO (QCD+EW)$\otimes$PS} & 389.3(8) & 393.1(8) & 1052(3) & 1066(3) & 766(2) & 774(2) \\
\hline
\end{tabular}
\caption{Total cross section results (in pb) of {\tt POWHEG-W\_EW} for
  $W^\pm\to \mu^\pm\nu_\mu$ production at the Tevatron ($\sqrt{S}=1.96$~TeV)
  and the LHC ($\sqrt{S}=7$~TeV) with bare acceptance criteria as listed in
  Section~\ref{sec:setup}. Shown are results for the LO, NLO EW, NLO QCD, the
  combined NLO QCD and EW cross sections, as well as results including
  showering performed with {\tt Pythia} or {\tt Herwig} as provided by the
  {\tt POWHEG Box}. The errors shown in parenthesis are statistical errors of
  the Monte Carlo integration. As a cross check the results of {\tt WGRAD2} are
  provided as well (in square brackets). }
\label{tab:totcsbare}
\end{center}
\end{table}

\begin{table}[H]
\begin{center}
\begin{tabular}{c|c|c|c|c|c|c|}
\cline{2-7} 
& \multicolumn{2}{|c|}{Tevatron} & \multicolumn{4}{|c|}{LHC}\\
\cline{2-7}
& \multicolumn{2}{|c|}{$W^+$} & \multicolumn{2}{|c|}{$W^+$} &\multicolumn{2}{|c|}{$W^-$}\\
\hline
\multicolumn{1}{|c|}{LO} & \multicolumn{2}{|c}{ 320.66(2) } &
\multicolumn{2}{|c|}{985.62(9) }& \multicolumn{2}{c|}{710.52(6) } \\   
\hline
\multicolumn{1}{|c|}{NLO EW} & \multicolumn{2}{|c|}{ 325.9(1)} & \multicolumn{2}{|c|}{999.0(5) }&
\multicolumn{2}{c|}{715.3(8)}\\   
\hline
\multicolumn{1}{|c|}{NLO QCD} & \multicolumn{2}{|c|}{369.75(4)} & \multicolumn{2}{|c|}{1037.9(2) }& \multicolumn{2}{|c|}{758.1(1) } \\   
\hline
\multicolumn{1}{|c|}{NLO (QCD+EW)} & \multicolumn{2}{|c|}{373.9(1) } & \multicolumn{2}{|c|}{1049.0(6) }& \multicolumn{2}{|c|}{765(1)} \\   
\hline
& \texttt{Pythia} & \texttt{Herwig}  & \texttt{Pythia} & 
\texttt{Herwig} & \texttt{Pythia} & \texttt{Herwig} \\
\hline
\multicolumn{1}{|c|}{LO$\otimes$PS} & 296.3(6) & 298.2(6) & 853(3) & 861(3) & 622(2) & 626(2) \\
\hline
\multicolumn{1}{|c|}{NLO QCD$\otimes$PS} & 358.4(8) & 361.4(8) & 1006(3) & 1019(3) & 736(2) & 743(2) \\
\hline
\multicolumn{1}{|c|}{NLO (QCD+EW)$\otimes$PS} & 364.1(8) & 366.9(8) & 1015(3) & 1026(3) & 743(2) & 751(2) \\
\hline
\end{tabular}
\caption{Total cross section results (in pb) of {\tt POWHEG-W\_EW} for $W^\pm\to \mu^\pm\nu_\mu$ production at the 
Tevatron ($\sqrt{S}=1.96$~TeV) and the LHC ($\sqrt{S}=7$~TeV) with calometric acceptance criteria 
as listed in Section~\ref{sec:setup}. Shown are results for LO, NLO EW, NLO
QCD, the combined NLO QCD and EW cross sections, as well as results including
showering performed with {\tt Pythia} or {\tt Herwig} as provided by the
{\tt POWHEG Box}. The errors shown in
parenthesis are statistical errors of the Monte Carlo integration. }
\label{tab:totcscalo}
\end{center}
\end{table}

\subsection{Transverse $W$ mass and charged lepton momentum distributions}\label{sec:distr}
\begin{figure}[H]
\begin{center}
\includegraphics[scale=0.8]{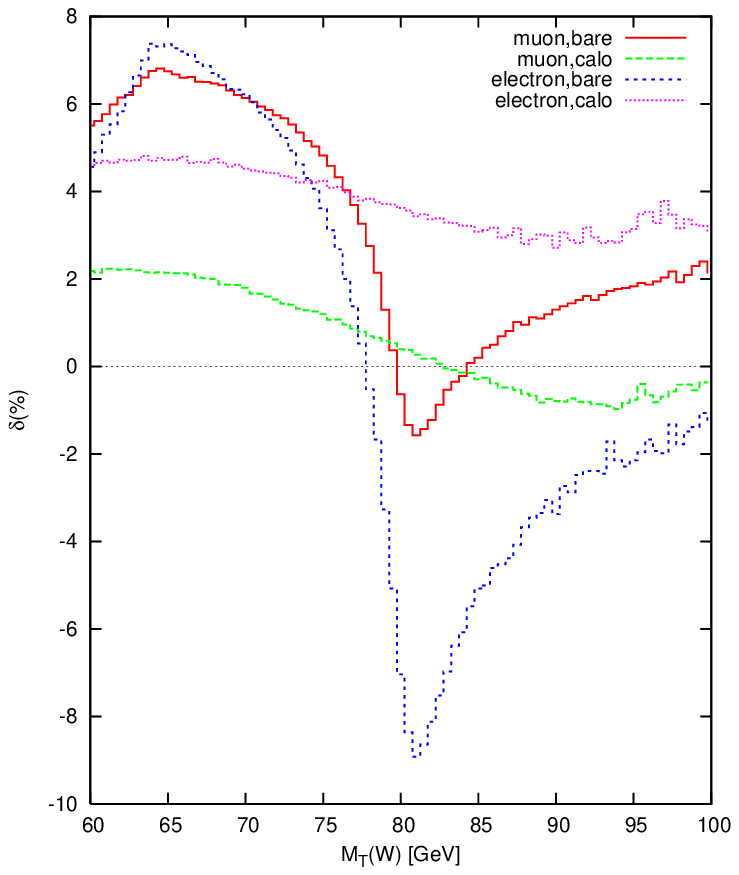}
\includegraphics[scale=0.8]{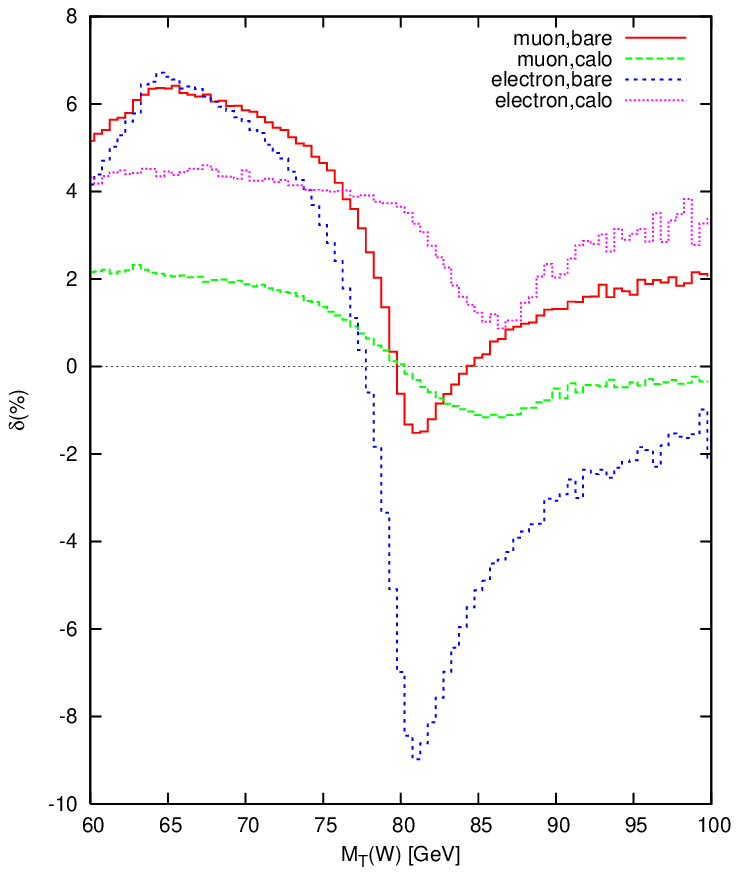}
\caption{Relative corrections $\delta_{\text{EW}}$ to the $M_T(W)$
  distributions for $p\bar{p}\to W^+\to e^+\nu_e,\mu^+\nu_\mu$ at
  $\sqrt{S}=1.96$ TeV on the left and $pp\to W^+\to e^+\nu_e,\mu^+\nu_\mu$ at
  $\sqrt{S}=7$ TeV on the right obtained with {\tt WGRAD2}. Bare and
  calometric cuts are shown for comparison.}
\label{fig:mtwdist1}
\end{center}
\end{figure}

\begin{figure}[H]
\begin{center}
\includegraphics[scale=0.8]{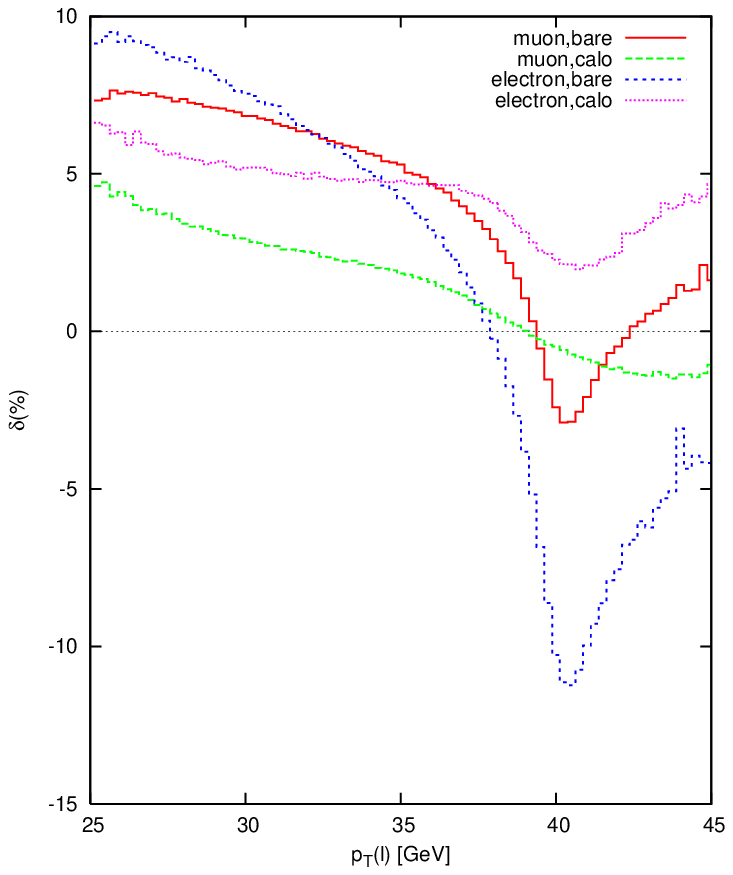}
\includegraphics[scale=0.8]{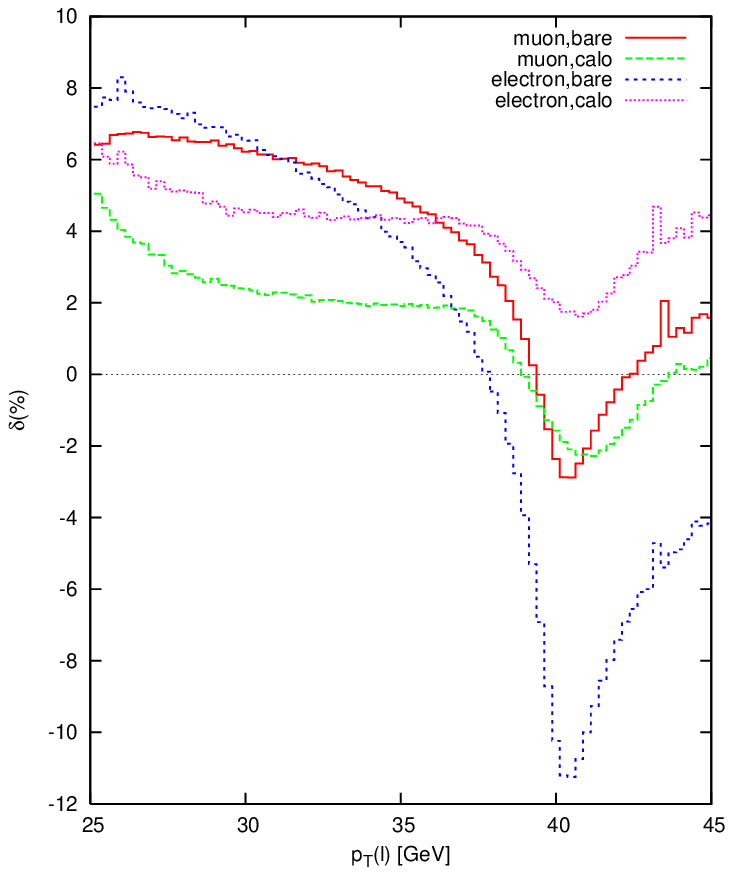}
\caption{Relative corrections $\delta_{\text{EW}}$ to $p_T(l)$ distributions
  for $p\bar{p}\to W^+\to e^+\nu_e,\mu^+\nu_\mu$ at $\sqrt{S}=1.96$ TeV, on
  the left and $pp\to W^+\to e^+\nu_e,\mu^+\nu_\mu$ at $\sqrt{S}=7$ TeV on the
  right obtained with {\tt WGRAD2}. Bare and calometric cuts are shown for
  comparison.}
\label{fig:ptldist1}
\end{center}
\end{figure}

\begin{figure}[H]
\begin{center}
\includegraphics[scale=1.0]{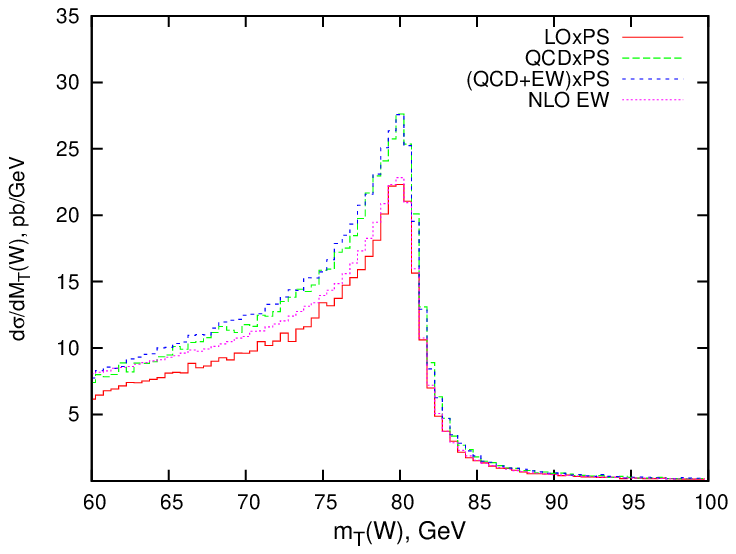}
\includegraphics[scale=1.0]{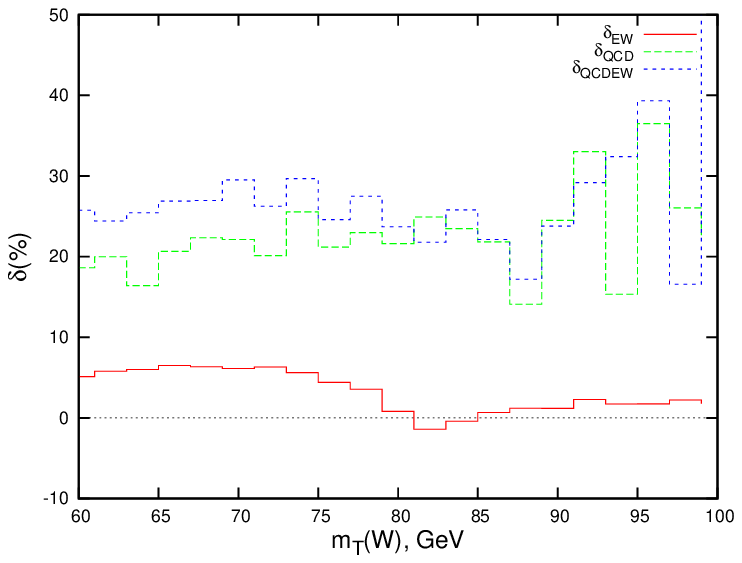}
\caption{$M_T(W)$ distributions and relative corrections
  $\delta_{\text{EW}},\delta_{\text{QCD}},\delta_{\text{QCDEW}}$ for
  $p\bar{p}\to W^+\to \mu^+\nu_\mu$, $\sqrt{S}=1.96$ TeV, obtained with {\tt
    POWHEG-W\_EW}, with bare cuts. Parton showering (denoted by PS) is
  performed by interfacing with {\tt Pythia}.}
\label{fig:mtwdist3}
\end{center}
\end{figure}

\begin{figure}[H]
\begin{center}
\includegraphics[scale=1.0]{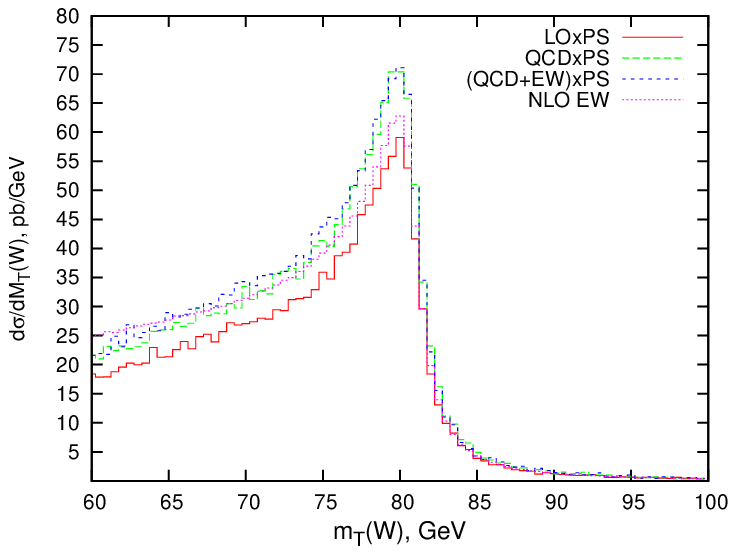}
\includegraphics[scale=1.0]{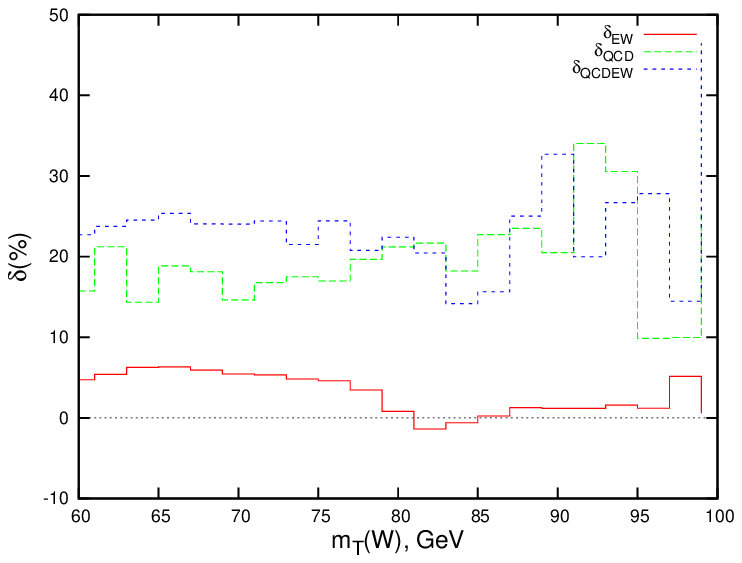}
\caption{$M_T(W)$ distributions and relative corrections
  $\delta_{\text{EW}},\delta_{\text{QCD}},\delta_{\text{QCDEW}}$ for $pp\to
  W^+\to \mu^+\nu_\mu$, $\sqrt{S}=7$ TeV, obtained with {\tt POWHEG-W\_EW},
  with bare cuts. Parton showering (denoted by PS) is performed by interfacing
  with {\tt Pythia}.}
\label{fig:mtwdist5}
\end{center}
\end{figure}

\begin{figure}[H]
\begin{center}
\includegraphics[scale=1.0]{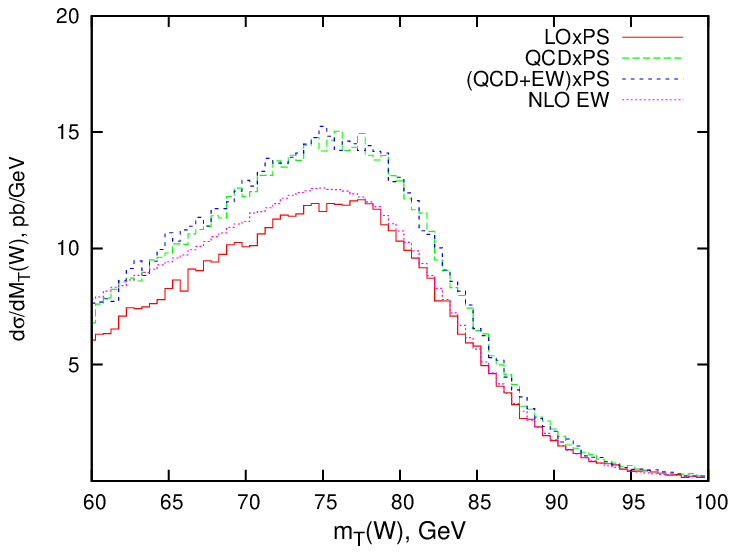}
\includegraphics[scale=1.0]{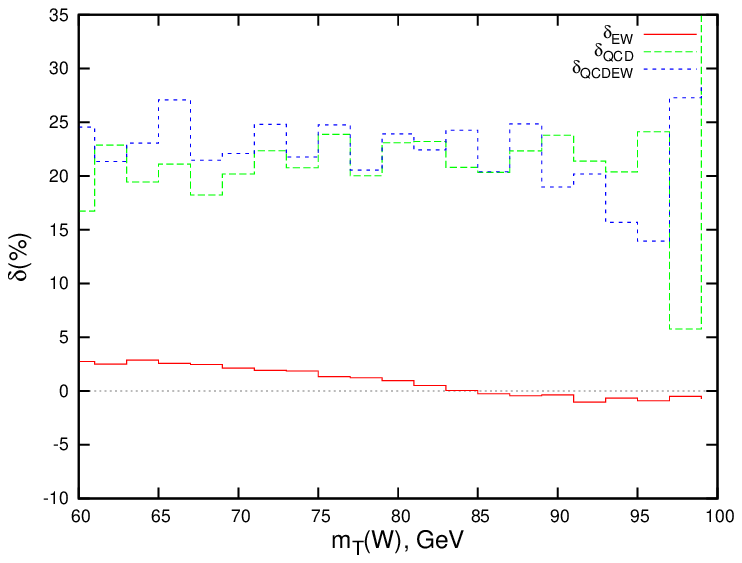}
\caption{$M_T(W)$ distributions and relative corrections
  $\delta_{\text{EW}},\delta_{\text{QCD}},\delta_{\text{QCDEW}}$ for
  $p\bar{p}\to W^+\to \mu^+\nu_\mu$, $\sqrt{S}=1.96$ TeV, obtained with {\tt
    POWHEG-W\_EW}, with calometric cuts. Parton showering (denoted by PS) is
  performed by interfacing with {\tt Pythia}.}
\label{fig:mtwdist7}
\end{center}
\end{figure}

\begin{figure}[H]
\begin{center}
\includegraphics[scale=1.0]{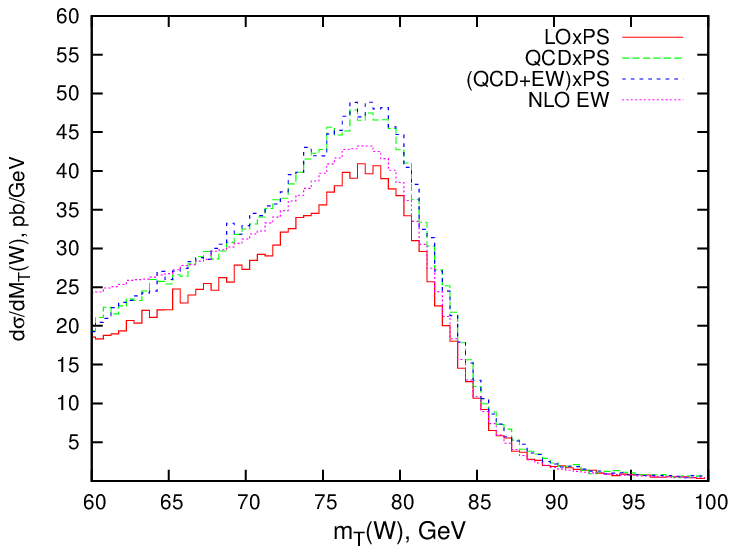}
\includegraphics[scale=1.0]{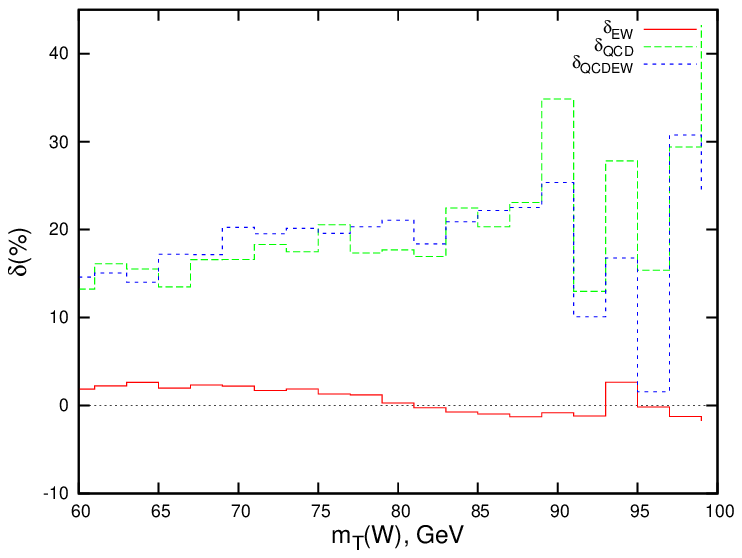}
\caption{$M_T(W)$ distributions and relative corrections
  $\delta_{\text{EW}},\delta_{\text{QCD}},\delta_{\text{QCDEW}}$ for $pp\to
  W^+\to \mu^+\nu_\mu$, $\sqrt{S}=7$ TeV, obtained with {\tt POWHEG-W\_EW},
  with calometric cuts. Parton showering (denoted by PS) is performed by
  interfacing with {\tt Pythia}.}
\label{fig:mtwdist9}
\end{center}
\end{figure}

\begin{figure}[H]
\begin{center}
\includegraphics[scale=1.0]{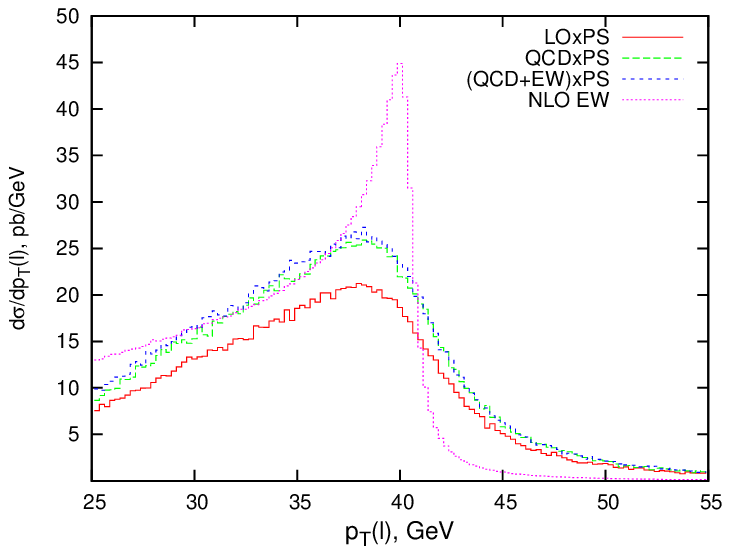}
\includegraphics[scale=1.0]{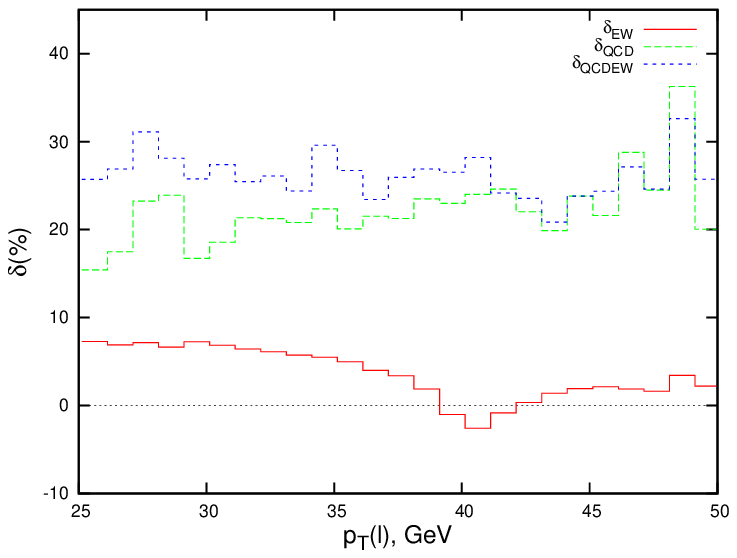}
\caption{$P_T(\mu)$ distributions and relative corrections $\delta_{\text{EW}},\delta_{\text{QCD}},\delta_{\text{QCDEW}}$ for 
$p\bar{p}\to W^+\to \mu^+\nu_\mu$, $\sqrt{S}=1.96$ TeV,
obtained with {\tt POWHEG-W\_EW}, with bare cuts. Parton showering (denoted by PS) is performed by interfacing with {\tt Pythia}.}
\label{fig:ptldist3}
\end{center}
\end{figure}

\begin{figure}[H]
\begin{center}
\includegraphics[scale=1.0]{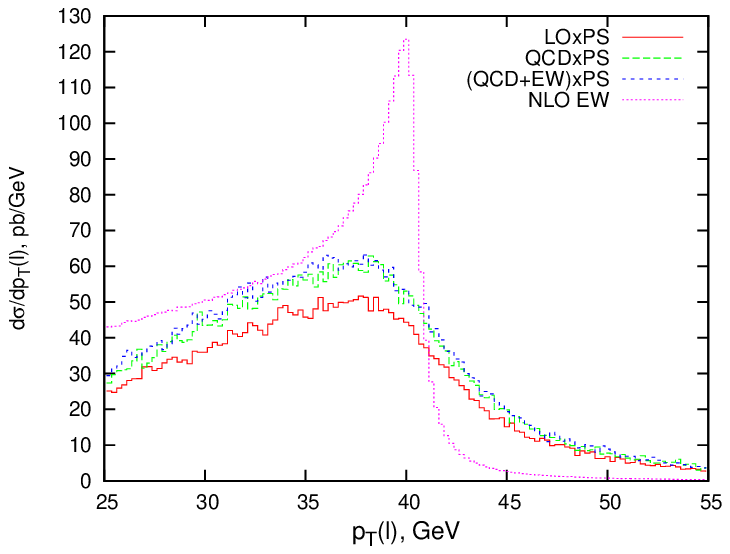}
\includegraphics[scale=1.0]{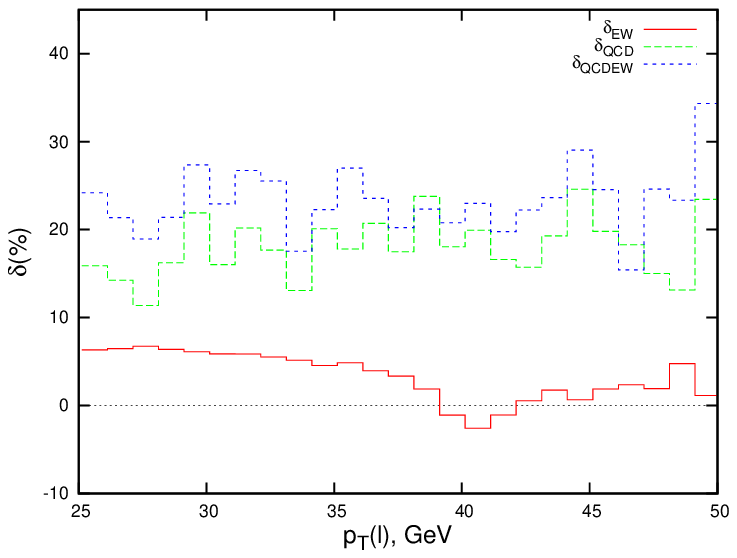}
\caption{$P_T(\mu)$ distributions and relative corrections $\delta_{\text{EW}},\delta_{\text{QCD}},\delta_{\text{QCDEW}}$ for 
$pp\to W^+\to \mu^+\nu_\mu$, $\sqrt{S}=7$ TeV,
obtained with {\tt POWHEG-W\_EW}, with bare cuts. Parton showering (denoted by PS) is performed by interfacing with {\tt Pythia}.}
\label{fig:ptldist5}
\end{center}
\end{figure}

\begin{figure}[H]
\begin{center}
\includegraphics[scale=1.0]{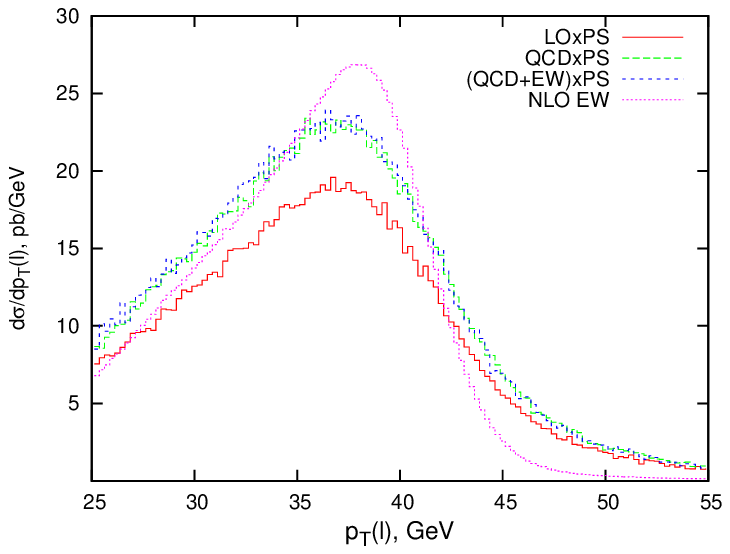}
\includegraphics[scale=1.0]{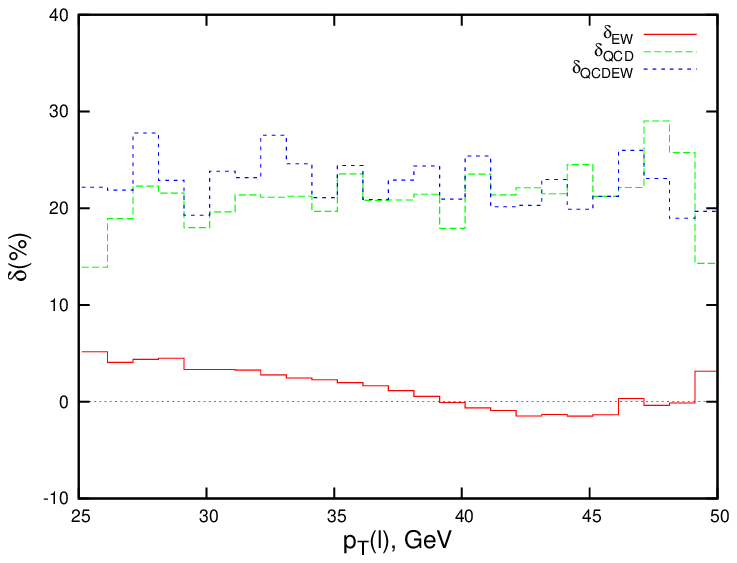}
\caption{$P_T(\mu)$ distributions and relative corrections $\delta_{\text{EW}},\delta_{\text{QCD}},\delta_{\text{QCDEW}}$ for 
$p\bar{p}\to W^+\to \mu^+\nu_\mu$, $\sqrt{S}=1.96$ TeV,
obtained with {\tt POWHEG-W\_EW}, with calometric cuts. Parton showering (denoted by PS) is performed by interfacing with {\tt Pythia}.}
\label{fig:ptldist7}
\end{center}
\end{figure}

\begin{figure}[H]
\begin{center}
\includegraphics[scale=1.0]{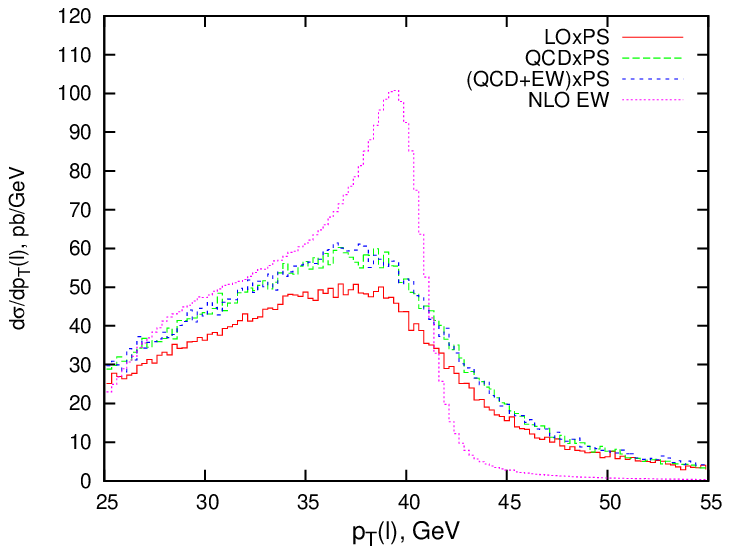}
\includegraphics[scale=1.0]{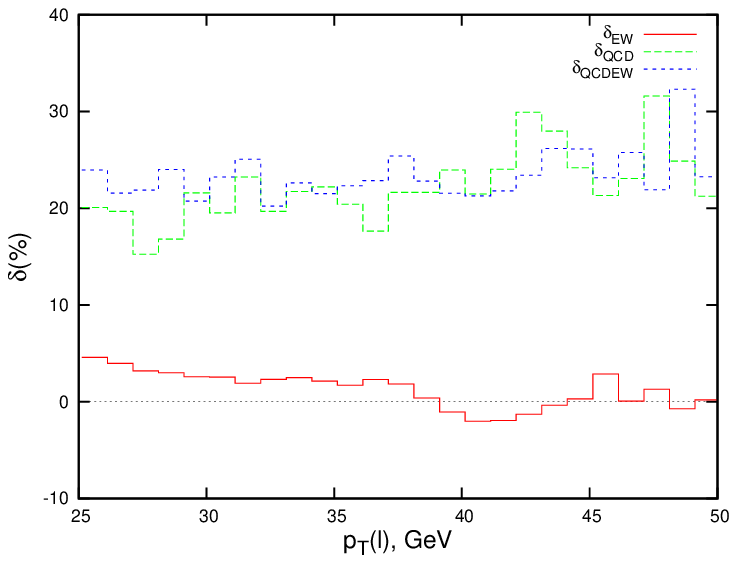}
\caption{$P_T(\mu)$ distributions and relative corrections $\delta_{\text{EW}},\delta_{\text{QCD}},\delta_{\text{QCDEW}}$ for 
$pp\to W^+\to \mu^+\nu_\mu$, $\sqrt{S}=7$ TeV,
obtained with {\tt POWHEG-W\_EW}, with calometric cuts. Parton showering (denoted by PS) is performed by interfacing with {\tt Pythia}.}
\label{fig:ptldist9}
\end{center}
\end{figure}

Differential distributions for the following $W$-boson observables are shown:
transverse mass of the $W$, $M_T(W)$, and the transverse momentum of the muon,
$p_{T}(\mu)$. The transverse $W$ mass is defined in terms of lepton-neutrino
pair observables as
\begin{equation}
M_T(W)=\sqrt{2p_{T}(\ell)p_{T}(\nu)(1-\cos(\Delta \phi_{\ell\nu}))}
\end{equation}
with $\phi$ the azimuthal angle of the charged lepton or neutrino and $\Delta
\phi_{\ell\nu}$ the difference between them.  Both observables are being used
to perform a high precision $W$ mass measurement at the
Tevatron~\cite{Aaltonen:2007ps,Abazov:2009cp}.  The $W$ mass extracted from
these observables is especially sensitive to changes in the lineshape in the
vicinity of the Jacobian peak. As has been well-studied in the literature,
final-state photon radiation greatly affects the distributions in this region
and predictions for these effects need to be under good theoretical control.
Here we will not provide a detailed phenomenological study of these EW
effects, which are available in the literature (see, e.g.,
Ref.~\cite{Gerber:2007xk} for an overview), but rather shall explore how the
impact of EW ${\cal O}(\alpha)$ corrections is affected by the presence of QCD
radiation when considering realistic lepton identification criteria. We only
briefly illustrate the main features of the impact of the EW ${\cal
  O}(\alpha)$ corrections on the $M_T(W)$ distribution in
Fig.~\ref{fig:mtwdist1} and on the $p_T(l=\mu,e)$ distribution in
Fig.~\ref{fig:ptldist1} when considering the bare and calometric setup by
showing the relative corrections defined as
\begin{equation}\label{eq:relew}
\delta_{\text{EW}}(\%)= 
\frac{  \frac{d\sigma_{\text{EW}}}{d\cal{O}} - \frac{d\sigma_{\text{LO}}}{d\cal{O}} } 
{\frac{d\sigma_{\text{LO}}}{d\cal{O}}} \times 100\,.
\end{equation}
The large distortion of the Jacobian peak, especially when only bare cuts are
applied, is due to collinear final-state photon radiation which results into
large logarithmic enhancements of the form $\alpha \log(m_l^2/\hat s)$, where
$m_l$ denotes the charged lepton mass and $\hat s$ the partonic CM energy
squared. In the electron case, when more realistic experimental conditions are
taken into account, the electron and photon four-momentum vectors are
recombined to an effective electron four-momentum vector if their separation
$\Delta R_{e\gamma}$ in the azimuthal angle--pseudorapidity plane is smaller
than a critical value $R_{cut}=0.1$. In that case, these mass-singular
logarithms cancel, and only a small effect of the EW corrections survives.  In
the muon case, however, the muon is well separated from the photon so that in
the calometric setup, the distortion of the $M_T(W)$ and $p_T(\mu)$
distributions around the Jacobian peak is more pronounced than in the electron
case. Since the electron case is very sensitive to the details of the lepton
identification requirements, a detailed study of combined EW and QCD effects
should be performed in collaboration with experimentalists involved in the $W$
mass measurement.  We leave such a study to a later publication and
concentrate in this paper on discussing the impact of QCD corrections on the
EW effects in the muon case.  Note that the results presented in
Ref.~\cite{Cao:2004yy} are obtained for the electron case with bare cuts, and
therefore larger effects have been observed than it can be the case with a
more realistic treatment.  Finally, we will only present results for the $W^+
\to \mu^+ \nu_\mu$ process, since even at the LHC the EW effects in the
$M_T(W)$ and $p_T(\mu)$ distributions are similar in $W^+$ and $W^-$
production, at least in the kinematic region of interest (note that there is
no distinction between $W^+$ and $W^-$ production at the Tevatron because of
the symmetric initial state).  Also in the presence of QCD radiation we found
that the relative corrections discussed in this paper exhibit very similar
features for $W^+$ and $W^-$ observables at the LHC.
In Figs.~\ref{fig:mtwdist3},~\ref{fig:mtwdist5},~\ref{fig:mtwdist7},~\ref{fig:mtwdist9}
and Figs.~\ref{fig:ptldist3},~\ref{fig:ptldist5},~\ref{fig:ptldist7},~\ref{fig:ptldist9}
we provide respectively $M_T(W)$ and $p_T(\mu)$ distributions calculated at
LO$\otimes$PS, NLO EW, NLO QCD$\otimes$PS and NLO (QCD+EW)$\otimes$PS as provided by {\tt
  POWHEG-W\_EW}, where we used {\tt Pythia} to shower the events.  We also
produced these distributions with {\tt Herwig}, but since the effects we are
interested in, i.~e. the change of the relative impact of EW corrections in the
presence of QCD radiation are similar we only show results obtained with {\tt
  Pythia}.  We note that a detailed, tuned comparison of the impact of these
two parton-shower MCs on $W$ observables in the presence of the complete EW
${\cal O}(\alpha)$ corrections, which is especially important in determining
their contribution to the theoretical uncertainty in the extraction of the $W$
mass, can now be conveniently performed with {\tt POWHEG-W\_EW}.  To
illustrate the impact of the various higher-order corrections on $M_T(W)$ and
$p_T(\mu)$ distributions we also show various relative corrections. Apart from
$\delta_{\text{EW}}$ of Eq.~\ref{eq:relew}, we show the impact of QCD
corrections ($\delta_{\text{QCD}}$) and combined QCD and EW corrections
($\delta_{\text{QCDEW}}$) in the presence of a QCD parton shower relative to
the LO parton shower result, defined as
\begin{equation}\label{eq:dqcd}
\delta_\text{QCD}(\%)= 
\frac{\frac{d\sigma_{{\text{QCD}}\otimes {\text PS}}}{d\cal{O}}-
\frac{d\sigma_{\text{LO}\otimes {\text PS}}}{d\cal{O}}}
{\frac{d\sigma_{\text{LO}\otimes {\text PS}}}{d\cal{O}}} \times 100\,,
\end{equation}
and
\begin{equation}\label{eq:dqcdew}
\delta_\text{QCDEW}(\%)= 
\frac{\frac{d\sigma_{{(\text{QCD+EW})}\otimes {\text PS}}}{d\cal{O}}-
\frac{d\sigma_{\text{LO}\otimes {\text PS}}}{d\cal{O}}}
{\frac{d\sigma_{\text{LO}\otimes {\text PS}}}{d\cal{O}}} \times 100\,,
\end{equation}
Note that the EW NLO results shown in
Figs.~\ref{fig:mtwdist3},~\ref{fig:mtwdist5},~\ref{fig:mtwdist7},~\ref{fig:mtwdist9}
and
Figs.~\ref{fig:ptldist3},~\ref{fig:ptldist5},~\ref{fig:ptldist7},~\ref{fig:ptldist9}
can be directly compared to and should agree with the results obtained with
{\tt WGRAD2} shown in Figs.~\ref{fig:mtwdist1} and~\ref{fig:ptldist1}.
%
Overall, we observe that the impact on the $M_T(W)$ distribution of the EW
corrections, $\delta_{\text{EW}}$, alone is seen to be slightly negative in
the peak region, while the effect of the QCD (NLO and PS),
$\delta_{\text{QCD}}$, corrections alone consistently increases the
LO$\otimes$PS $M_T(W)$ distribution, as expected.  Now, the combined relative
NLO (QCD+EW)$\otimes$PS corrections, $\delta_{\text{QCDEW}}$, still follow in
magnitude the effect one expects when simply adding NLO QCD$\otimes$PS and NLO
EW corrections, but seems to be slightly different in shape in the peak region
at the LHC when applying bare cuts. The dip observed in the relative
$\delta_{\text{EW}}$ correction seems now to be somewhat washed out.

\begin{figure}[H]
\begin{center}
\includegraphics[scale=1.0]{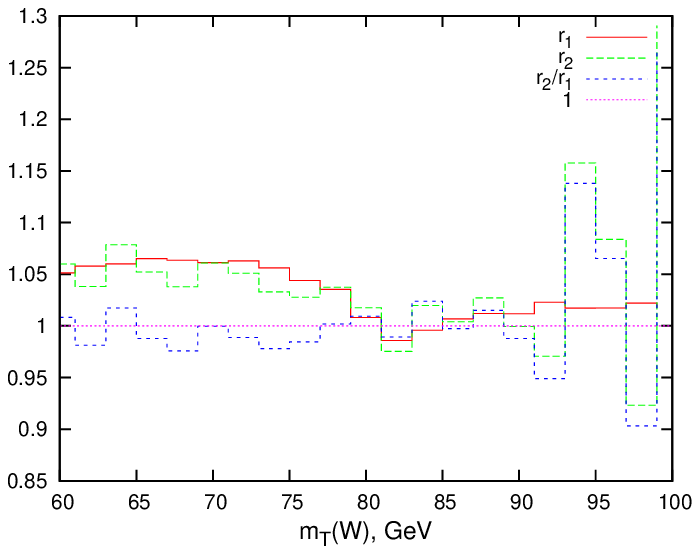}
\includegraphics[scale=1.0]{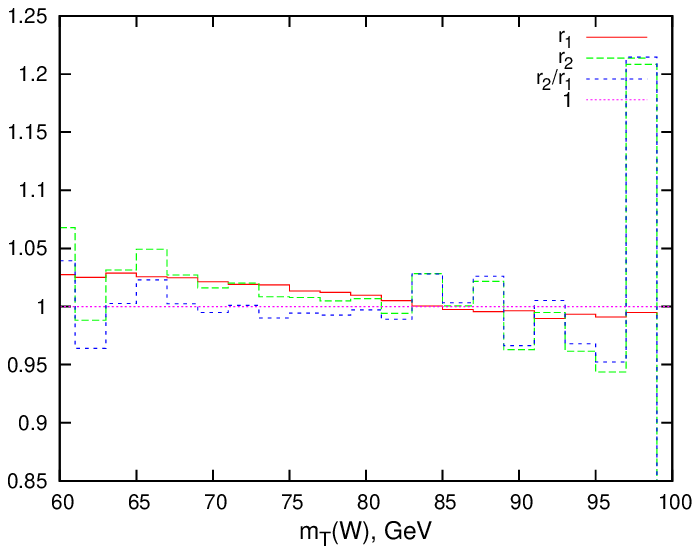}
\caption{Ratios $r_{1,2}$ and their ratio ${\cal R}$ for $p\bar{p}\to W^+\to \mu^+\nu_\mu$, $\sqrt{S}=1.96$ TeV,
obtained with {\tt POWHEG-W\_EW}, with bare (left) and calometric (right) cuts. Parton showering is performed by interfacing with {\tt Pythia}.}
\label{fig:mtwdist4}
\end{center}
\end{figure}

\begin{figure}[H]
\begin{center}
\includegraphics[scale=1.0]{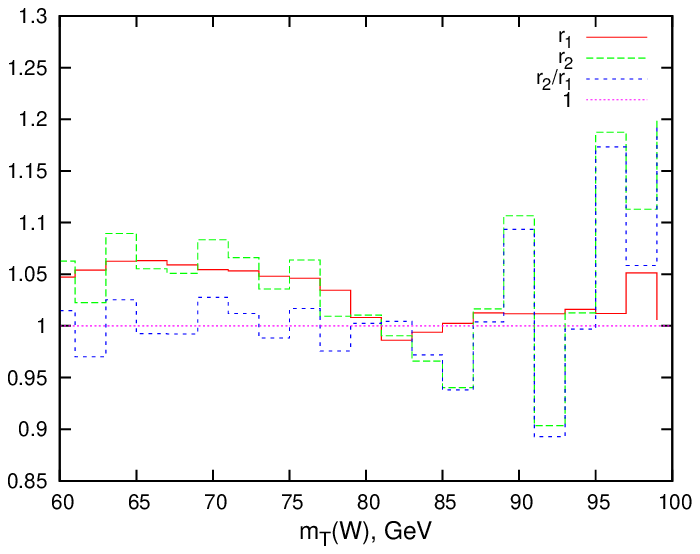}
\includegraphics[scale=1.0]{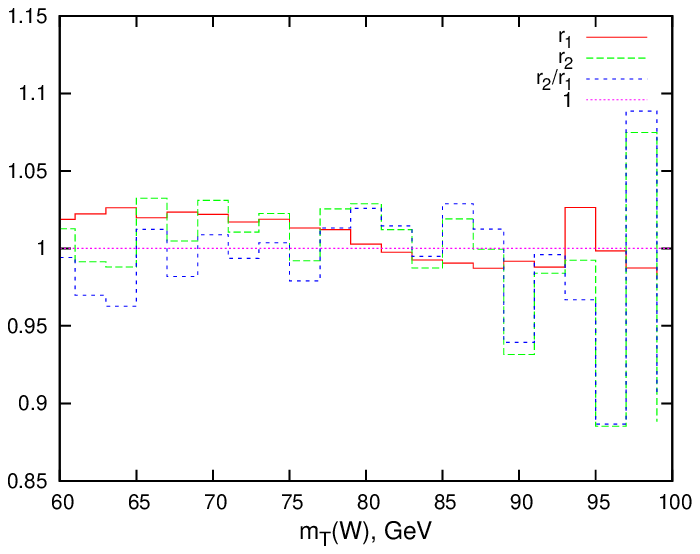}
\caption{Ratios $r_{1,2}$ and their ratio ${\cal R}$ for $pp\to W^+\to \mu^+\nu_\mu$, $\sqrt{S}=7$ TeV,
obtained with {\tt POWHEG-W\_EW}, with bare (left) and calometric (right) cuts. Parton showering is performed by interfacing with {\tt Pythia}.}
\label{fig:mtwdist10}
\end{center}
\end{figure}

To study this possible effect of combining EW and QCD corrections more
closely, we follow the discussion of Ref.~\cite{Cao:2004yy}, and for each of
the observables, we define
\begin{eqnarray}\label{eq:ratios}
r_1&=& 
\frac{\frac{d\sigma_{\text{NLO EW}}}{d\cal{O}}}
{\frac{d\sigma_{\text{LO}}}{d\cal{O}}}\\ \nonumber
r_2&=& 
\frac{\frac{d\sigma_{(\text{QCD+EW})\otimes \text{PS}}}{d\cal{O}}}
{\frac{d\sigma_{\text{QCD}\otimes \text{PS}}}{d\cal{O}}} 
\end{eqnarray}
and show in Figs.~\ref{fig:mtwdist4},~\ref{fig:mtwdist10} $r_{1,2}$ together
with their ratio ${\cal R}=r_2/r_1$.  We see that within the statistical
uncertainty of the numerical integration, ${\cal R}$ is largely consistent with
unity, i.~e. generally the NLO EW corrections in the presence of QCD effects
behave like those of EW alone, only at the LHC there seems to be a slight change
in shape above the peak region, but given the large fluctuation in this region
this is more likely a relic of the numerical integration.  Since QCD radiation is
known not to have a significant effect on the shape of the $M_T(W)$
distribution, it is no surprise that the main features of the EW corrections
in the presence of QCD corrections are largely unchanged. This has also been
observed in Refs.~\cite{Cao:2004yy,Balossini:2009sa}. Our results
cannot be directly compared with earlier studies such as those presented in
Refs.~\cite{Cao:2004yy,Balossini:2009sa}, since in Ref.~\cite{Cao:2004yy} 
results are provided for the electron case with bare cuts and in
Ref.~\cite{Balossini:2009sa} slightly different selection criteria have been
used and the $G_\mu$ EW input scheme has been adopted~\footnote{The 
results obtained in the $G_\mu$ scheme can be estimated from our results in
the $\alpha(0)$ scheme by $\sigma_{G_\mu}^{NLO}=
(\sigma_\alpha^{NLO}-2\Delta \, r \sigma^{LO}_\alpha)
\alpha_{G_\mu}^2/\alpha(0)^2$ with $\Delta \, r = 0.0315$, $\alpha_{G_\mu}=
\sqrt{2} G_\mu \sin^2\theta_w M_W^2/\pi$, and $G_\mu=1.16639 \cdot
10^{-5}$~${\rm GeV}^{-2}$.}. Nevertheless, the overall
features of the effects of combining QCD and EW corrections turn out to be
similar. It will be interesting to perform a tuned comparison of different
implementations of EW/QED corrections in NLO QCD+resummed calculations of $W$
boson observables, i.~e. of {\tt POWHEG-W\_EW}, {\tt
  MC@NLO/HORACE}~\cite{Balossini:2009sa} and {\tt ResBos-A}~\cite{Cao:2004yy},
which is left to a future publication.

We now turn to the discussion of the $p_T(\mu)$ distributions, and show in the right-hand plots in
Figs.~\ref{fig:ptldist3}-\ref{fig:ptldist9} again the relative corrections  
$\delta_{\text{EW}}$ of Eq.~\ref{eq:relew}, $\delta_{\text{QCD}}$ of Eq.~\ref{eq:dqcd} and 
$\delta_{\text{QCDEW}}$ of Eq.~\ref{eq:dqcdew}. In contrast to the $M_T(W)$
distribution, now the effects of QCD radiation on the shape of the $p_T(\mu)$ distributions are quite pronounced, as can be
seen by comparing the exact NLO EW distribution with the distributions obtained
with showered events shown on the left-hand side of Figs.~\ref{fig:ptldist3}-\ref{fig:ptldist9}. 
So, we expect to see a change in how the EW corrections impact the $p_T(\mu)$
distributions due to a non-trivial interplay of QCD and EW corrections in the
combined result.  After all, as can be seen in Eq.~\ref{eq:nason} with $\bar
B$ of Eq.~\ref{eq:main}, higher-order
QCD-EW interference terms are present and may have a non-negligible impact on
the $p_T(\mu)$ distributions. Therefore, it is interesting to note the
overall significant shape change between the exact NLO EW and the combined 
NLO (QCD+EW)$\otimes$PS $p_T(\mu)$ distributions.  That the NLO EW corrections, in the
presence of QCD effects, tend to be quite subdued is obvious in the magnitude
and shape of the relative NLO (QCD+EW)$\otimes$PS corrections compared to the
NLO EW relative corrections, as shown on
the right-hand side of these figures.  Especially the dip in the Jacobian region from the
NLO EW corrections with bare cuts is now mostly washed out.

\begin{figure}[H]
\begin{center}
\includegraphics[scale=1.0]{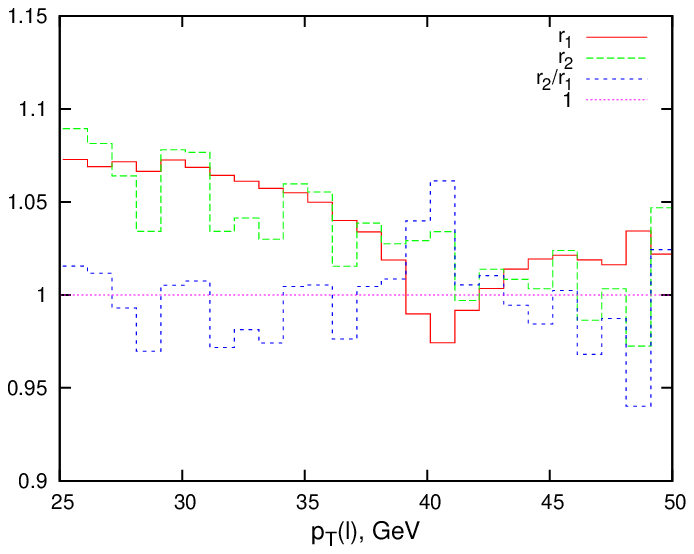}
\includegraphics[scale=1.0]{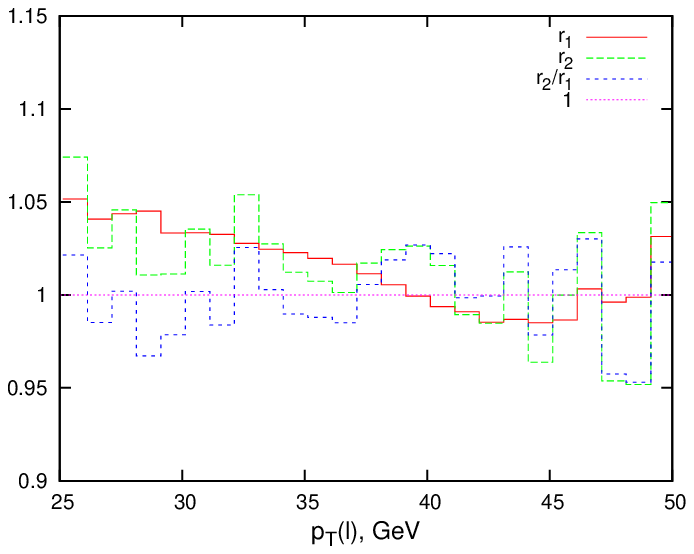}
\caption{Ratios $r_{1,2}$ and their ratio ${\cal R}$ for $p\bar{p}\to W^+\to \mu^+\nu_\mu$, $\sqrt{S}=1.96$ TeV,
obtained with {\tt POWHEG-W\_EW}, with bare (left) and calometric (right) cuts. Parton showering is performed by interfacing with {\tt Pythia}.}
\label{fig:ptldist4}
\end{center}
\end{figure}

\begin{figure}[H]
\begin{center}
\includegraphics[scale=1.0]{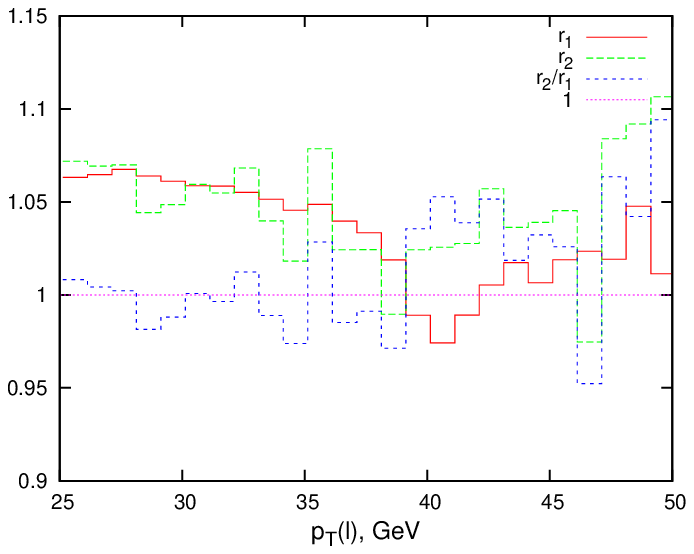}
\includegraphics[scale=1.0]{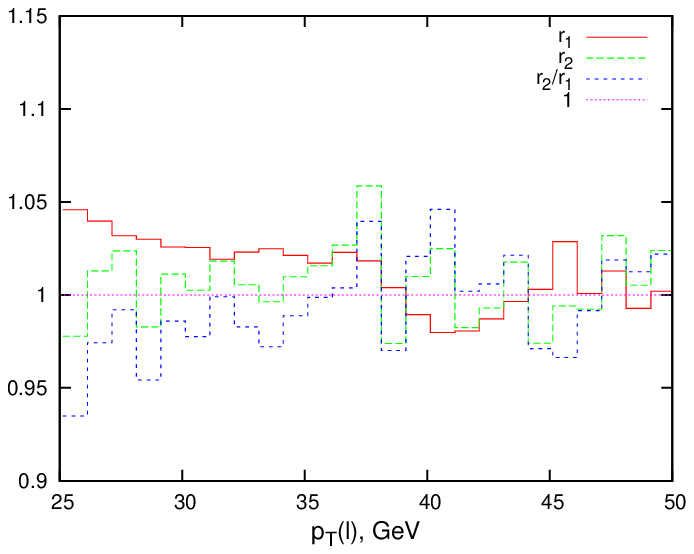}
\caption{Ratios $r_{1,2}$ and their ratio ${\cal R}$ for $pp\to W^+\to \mu^+\nu_\mu$, $\sqrt{S}=7$ TeV,
obtained with {\tt POWHEG-W\_EW}, with bare (left) and calometric (right) cuts. Parton showering is performed by interfacing with {\tt Pythia}.}
\label{fig:ptldist10}
\end{center}
\end{figure}

Again, we study these effects more closely in Figs.~\ref{fig:ptldist4}
and~\ref{fig:ptldist10} by comparing the impact of NLO EW corrections alone
and their impact in the presence of QCD corrections as described by $r_1$ and
$r_2$ of Eq.~\ref{eq:ratios} respectively. Clearly, unlike in the case 
of the
$M_T(W)$ distribution, $r_1$ and $r_2$ are now quite different so that their
ratio ${\cal R}$ exhibits an interesting shape especially around the Jacobian
peak. The same feature has also been observed in the ${\cal R}$ ratio of Ref.~\cite{Cao:2004yy}. As expected, in the calometric setup, this difference is much less
pronounced due to the smaller impact of the EW corrections on the shape of the
$p_T(\mu)$ distributions.

\section{Conclusions}\label{sec:conclusions}

In this paper we described the combination of the complete EW ${\cal O}(\alpha)$
corrections with NLO QCD+resummed corrections to $W$ production in hadronic
collisions, based on implementing the EW corrections of {\tt WGRAD2} in {\tt
  POWHEG-W}. Using the resulting MC program, {\tt POWHEG-W\_EW}, which is
publicly available on the webpage of the {\tt POWHEG BOX}, we presented
results for the transverse $W$ mass and charged lepton momentum distributions,
taking into account lepton identification requirements which are closely modeled after
those used in the high-precision measurement of the $W$ mass at the Tevatron.
In view of the anticipated precision of the $W$ mass measurement at the Tevatron
and the LHC, predictions for these observables including higher-order
radiative corrections have to be under excellent control.  Tools such as {\tt
  POWHEG-W\_EW} that allow the study of combined EW and QCD corrections are
important in reducing the theoretical uncertainty in the $W$ mass measurement.
We especially concentrated on studying whether there is a change of the impact
of EW corrections when QCD radiation is present as described by {\tt
  POWHEG+Pythia} in the kinematic region where EW corrections are known to
have a significant impact on the extracted $W$ mass. We found interesting
QCD-EW interference effects in the $p_T(\mu)$ distributions, i.~e. effects
that go beyond simply adding QCD and NLO EW corrections, that change the shape
of the distribution around the Jacobian peak.  These effects are similar to
those observed in Refs.~\cite{Cao:2004yy,Balossini:2009sa}, and their impact
on the $W$ mass extracted from the $p_T(\mu)$ distribution should be studied
in more detail by using realistic detector resolution effects, ideally in 
close collaboration with the experimentalists performing the
$W$-mass measurement.  Moreover,
these findings also suggest that a calculation of the complete mixed EW-QCD
${\cal O}(\alpha \alpha_s)$ corrections is desirable to further reduce the
theoretical uncertainty and to obtain an accurate estimate of the theory
uncertainty due to missing higher-order corrections. Further improvements that
are planned for {\tt POWHEG-W\_EW} include the implementation of the known
higher-order QED and EW effects, i.~e. beyond NLO, of photon-induced
processes, and the usage of an updated PDF that fully considers QED
corrections, once available. Since {\tt POWHEG-W\_EW} interfaces to both {\tt
  Pythia} and {\tt Herwig} it is also a convenient tool to perform a tuned
comparison of QCD+EW effects when using either parton-shower MC.  Finally,
since {\tt POWHEG-W\_EW} includes the complete NLO EW corrections, it is
interesting to note that the effects of EW Sudakov logarithms that become
numerically important in distributions at high energies can now also be studied
in the presence of QCD radiation. This is especially interesting for the
search for $W'$ bosons at the LHC.

\newpage
\begin{center} {\bf Note Added} \end{center}

Shortly after the submission of our arXiv paper, another implementation of 
EW corrections to single $W$ production into the {\tt POWHEG Box}
became available~\cite{Barze:2012tt}. 

\section*{Acknowledgments}
We would like to thank Carlo Oleari and Paolo Nason for helpful discussions
and their support in making this addition to the {\tt POWHEG BOX} public.  We
are grateful to Ulrich Baur, who sadly passed away in November 2010, for his
invaluable feedback and support in the beginning stages of this project, This
research was supported in part by the National Science Foundation under grant
No.~NSF-PHY-0547564 and No.~NSF-PHY-0757691, and a LHC-TI fellowship (NSF
No.~PHY-0705682) and Verbundprojekt: ``QCD-Jets in Higgs-Physik
und Suchen nach neuer Physik'' (05H09VHE).  Part of this work was performed 
at the Institute for Theoretical Particle Physics at the Karlsruhe Institute 
of Technology (KIT), during D.~W.'s visit as a Mercator Fellow, funded by the Deutsche
Forschungsgemeinschaft (DFG). D.~W. is especially grateful to Johann K\"uhn
for his kind hospitality during her stay at the KIT.
 
\section{Appendix}
In the following we describe in detail the implementation of each EW piece
framed in Eq.~\ref{eq:main}, which are added to the QCD
contribution after testing if the user requests to include EW
corrections (\texttt{wgrad2=1}). An attempt is made to describe both the
analytical pieces in the code as well as their variable names.  This is
important so the user can be aware of how certain flags affect their values as
well as of which part of the EW corrections is contained in each piece. For
more explicit expressions of the EW ${\cal O}(\alpha)$ corrections see
Refs.\cite{Wackeroth:1996hz,Baur:1998kt,Baur:2004ig}.

\subsection{The virtual+soft finite piece, $V_{\textnormal{EW}}^{f_b}(\Phi_2)$}

The quantity $V_{\textnormal{EW}}^{f_b}(\Phi_2)$ is defined in
\texttt{POWHEG-BOX/W\_EW-BW/virtual\_EW.f}.  There are two flags which affect the outcome, both can
be set in the proper {\tt POWHEG} input file.  Except for
\texttt{qnonr}, all of the flags are used in the definitions of the
other boxed terms in Eq.~\ref{eq:main}.  For completeness, the
flags and their descriptions are listed below.
\begin{itemize}
\item[$\bullet$]{\texttt{QED}:
This flag toggles between different subsets of NLO QED contributions.} 
\begin{itemize}
\item[$\bullet$]{\texttt{QED=1} Initial State (IS) photon radiation. }
\item[$\bullet$]{\texttt{QED=2} Final State (FS) photon radiation. }
\item[$\bullet$]{\texttt{QED=3} interference between IS and FS photon radiation.}
\item[$\bullet$]{\texttt{QED=4} IS, FS and interference}
\end{itemize}
\end{itemize}
Note that the gauge invariant separation into IS and FS QED contributions has been performed 
according to Ref.~\cite{Wackeroth:1996hz}.

\begin{itemize}
\item[$\bullet$]{\texttt{qnonr}: 
This flag toggles between the treatment of EW corrections to resonant/non-resonant 
$W$ production. In the case of resonant $W$ production (qnonr=0), a gauge invariant separation 
into weak and QED $\mathcal{O}(\alpha)$ corrections and into IS and FS contributions is available according to Ref.~\cite{Wackeroth:1996hz}.
In this case, no weak box diagrams are included and the weak form factors are evaluated at
$\hat s=M_W^2$.}
\begin{itemize}
\item[$\bullet$]{\texttt{qnonr=0} excludes the weak box diagrams, corresponds to resonant $W$ 
production only.}
\item[$\bullet$]{\texttt{qnonr=1} includes the weak box diagrams, i.e. includes the complete EW 
$\mathcal{O}(\alpha)$ corrections to $W$ production.}
\end{itemize}
\end{itemize}
In choosing \texttt{qnonr=1}, the full set of virtual diagrams are used, including
the weak box diagrams which are necessary to describe non-resonant $W$ production.
Since in case of non-resonant $W$ production the gauge invariant separation in IS and FS
contributions according to Ref.~\cite{Wackeroth:1996hz} is no longer possible,
the entire set of real radiation diagrams must be included. Therefore, 
if choosing \texttt{qnonr=1} the user must also set \texttt{QED=4}. 

The contents of $V_{\textnormal{EW}}^{f_b}(\Phi_2)$, without the PDF factors, are now explicitly shown.
\begin{equation}
V_{\textnormal{EW}}^{f_b}(\Phi_2) \sim \texttt{soft(1:12) + virt(1:12)} \label{virtarrEW}
\end{equation}
The choice of \texttt{qnonr} affects \texttt{virt(1:12)} as follows.  For \texttt{qnonr=0}
\begin{equation}
\texttt{virt(1:12)} = \texttt{br\_born(1:12)}\mathcal{R}e\left[\texttt{fvwp(1)}_{\text{IS}}
+\texttt{fvwp(2)}_{\text{FS}}\right]\frac{2\pi}{\alpha_s}
\end{equation}
\texttt{br\_born(1:12)} are {\tt POWHEG-W}'s definition of the Born contribution and 
are equivalent to $\frac{1}{2\hat{s}}\overline{\sum}|M_B|^2$ for each flavor structure. 
\texttt{fvwp(1:2)} are the modified weak one-loop contributions for resonant $W$ production
and are defined in \texttt{/POWHEG-BOX/W\_EW-BW/libweak.f}.  
The user has the choice
to include separately the IS and FS contributions through the proper choice of 
\texttt{QED}.  Shown above is the outcome corresponding to \texttt{QED=4}.
For \texttt{qnonr=1}, 
\begin{align*}
\hspace{-9em}\texttt{virt(1:6)}&=|\texttt{CKM(1:6)}|^2\,
\frac{\pi^2\alpha^2}{s_W^4}\,\frac{1}{|\hat{s}-M_W^2+i\Gamma_WM_W|^2}
\frac{1}{2\hat{s}}\,\frac{1}{3}\,
\end{align*}
\begin{align}
\hspace{16em}\times\texttt{a2qqw(}\hat{s},-2k_{\bar{q}}\cdot k_l,-2k_q\cdot k_l, 16(k_{\bar{q}}\cdot 
k_{\bar{l}})^2)
\end{align}
\begin{align*}
\hspace{-9em}\texttt{virt(7:12)}&=|\texttt{CKM(7:12)}|^2\,
\frac{\pi^2\alpha^2}{s_W^4}\,\frac{1}{|\hat{s}-M_W^2+i\Gamma_WM_W|^2}
\frac{1}{2\hat{s}}\,\frac{1}{3}\,
\end{align*}
\begin{align}
\hspace{16em}\times\texttt{a2qqw(}\hat{s},-2k_q\cdot k_l,-2k_{\bar{q}}\cdot k_l,16(k_q\cdot k_{\bar{l}})^2)
\end{align}
The \texttt{function a2qqw(\ldots)} corresponds to the full weak one-loop 
contribution and is also defined in \texttt{/POWHEG-BOX/W\_EW-BW/libweak.f}.
The \texttt{soft} contributions of Eq.~\ref{virtarrEW} are given by
\begin{align*}
\hspace{-9em}\texttt{soft(1:6)}&=|\texttt{CKM(1:6)}|^2\,\frac{\pi^2\alpha^2}{s_W^4}\,
\frac{1}{2\hat{s}}\frac{16}{3}   
\frac{(k_{\bar{q}}\cdot k_{\bar{l}})^2}{|\hat{s}-M_W^2+i\Gamma_WM_W|^2}
\end{align*}
\begin{align}
\hspace{16em}\times\left[
\texttt{k\_fac(2)}_{\text{IS}} +\texttt{k\_fac(2)}_{\text{FS}} +\texttt{k\_fac(2)}_{\text{int}} 
\right]
\end{align}
\begin{align*}
\hspace{-9em}\texttt{soft(7:12)}&=|\texttt{CKM(7:12)}|^2\,\frac{\pi^2\alpha^2}{s_W^4}\,
\frac{1}{2\hat{s}}\frac{16}{3}
\frac{(k_q\cdot k_{\bar{l}})^2}{|\hat{s}-M_W^2+i\Gamma_WM_W|^2}
\end{align*}
\begin{align}
\hspace{16em}\times\left[
\texttt{k\_fac(1)}_{\text{IS}} +\texttt{k\_fac(1)}_{\text{FS}} +\texttt{k\_fac(1)}_{\text{int}} 
\right]
\end{align}
The \texttt{k\_fac(1:2)} terms correspond to the finite soft photon contributions and
are defined in \texttt{/POWHEG-BOX/W\_EW-BW/libqed.f}.  Again, the case of \texttt{QED=4} is
shown above, but the user can include any or all of these \texttt{k\_fac} terms by 
adjusting this flag. The \texttt{k\_fac}, and hence the soft contribution depends on the
soft PSS parameter, $\delta_s$, and when {\texttt collcut=1} is chosen also on $\delta_c$.

\subsection{The collinear remnants, $G^{1,f_b}_{\textnormal{EW},\oplus}$ , 
$G^{1,f_b}_{\textnormal{EW},\ominus}$}
These are defined in  \texttt{/POWHEG-BOX/W\_EW-BW/collinear\_EW.f}.
There is also the inclusion of a new flag, {\tt lfc}:  
\begin{itemize}
\item[$\bullet$]{\texttt{lfc}:
This flag controls the choice of the QED factorization scheme.}
\begin{itemize}
\item[$\bullet$]{\texttt{lfc=0}  $\overline{\text{MS}}$ scheme}
\item[$\bullet$]{\texttt{lfc=1}  DIS scheme}
\end{itemize} 
\end{itemize} 
{\tt POWHEG} only implements the QCD $\overline{\text{MS}}$ scheme.  Hence, as this flag
only applies to the EW portions, one has the option to choose different schemes for the 
QED and QCD factorization.  We now describe analytically the contents of these contributions.
\begin{equation*}
\hspace{-9.5em}
G^{1,1:6}_{\textnormal{EW},\oplus}+G^{1,1:6}_{\textnormal{EW},\ominus}
= |\texttt{CKM(1:6)}|^2\frac{\pi^2\alpha^2}{s_W^4}\,
\frac{1}{2\hat{s}}\frac{16}{3}\left(k_{\bar{q}} \cdot k_{\bar{l}}\right)^2
\frac{\texttt{kn\_jacborn}}{|\hat{s}-M_W^2+i\Gamma_WM_W|^2}
\end{equation*}
\begin{equation}
\hspace{0em}\times \left[\underbrace{\frac{1}{9}\bar{q}_1\left(\frac{x_1}{z_1},\mu_R\right)q_2(x_2,\mu_R)
\frac{\texttt{splitz1}}
{z_1}(1-x_1)}_{\sim\,G^1_{\oplus,\text{EW}}} +\underbrace{\frac{4}{9}\bar{q}_1(x_1,\mu_R)q_2\left(\frac{x_2}{z_2},\mu_R\right)
\frac{\texttt{splitz2}}{z_2}(1-x_2)}_{\sim\,G^1_{\ominus,\text{EW}}}\right]
\end{equation}
\vspace{-.25em}
\begin{equation*}
\hspace{-8.5em}
G^{1,7:12}_{\textnormal{EW},\oplus}+G^{1,7:12}_{\textnormal{EW},\ominus}
= |\texttt{CKM(7:12)}|^2\frac{\pi^2\alpha^2}{s_W^4}\,
\frac{1}{2\hat{s}}\frac{16}{3}\left(k_q \cdot k_{\bar{l}}\right)^2
\frac{\texttt{kn\_jacborn}}{|\hat{s}-M_W^2+i\Gamma_WM_W|^2}
\end{equation*}
\begin{equation}
\hspace{0em}\times \left[\underbrace{\frac{4}{9}q_1\left(\frac{x_1}{z_1},\mu_R\right)\bar{q}_2(x_2,\mu_R)
\frac{\texttt{splitz1}}
{z_1}(1-x_1)}_{\sim\,G^1_{\oplus,\text{EW}}}+\underbrace{\frac{1}{9}q_1(x_1,\mu_R)\bar{q}_2\left(\frac{x_2}{z_2},\mu_R\right)
\frac{\texttt{splitz2}}{z_2}(1-x_2)}_{\sim\,G^1_{\ominus,\text{EW}}}\right]
\end{equation}
\texttt{splitz1} and \texttt{splitz2} are functions of $z_1$ and $z_2$ respectively. For a generic $z$ they
are
\begin{equation}
\texttt{splitz} = \frac{\alpha}{2\pi}\left[\frac{1+z^2}{1-z}\ln\left(\frac{\hat{s}}{\mu_F^2}
\frac{z}{(1-z)^2}\frac{\delta_c}{2}\right)+1-z-\texttt{lfc}*\texttt{fcollz}\right]
\label{splitz}
\end{equation} 
with
\begin{equation}
\texttt{fcollz} = \frac{1+z^2}{1-z}\ln\left(\frac{1-z}{z}\right)-\frac{3}{2(1-z)}+2z+3
\end{equation} 
In Eq.~\ref{splitz} above, one sees explicitly the QED factorization scheme dependence. Note that the soft part of 
the QED PDF counterterm is included in \texttt{k\_fac}. Also, the dependence on the PSS parameter $\delta_c$
is explicitly shown. 

\subsection{The real contribution, $R_{\textnormal{EW}}^{f_b}$}
$R_{\textnormal{EW}}^{f_b}$ are defined in \texttt{/POWHEG-BOX/W\_EW-BW/real\_EW.f}.  
As mentioned earlier, {\tt POWHEG-W} considers not only 
$q \bar{q}$, but $gq$ and $g\bar{q}$ induced processes, while the calculation of 
{\tt WGRAD2} only includes $q \bar{q}$ induced processes.  This is why in 
Eq.~\ref{eq:main} there is only one $R_{\textnormal{EW}}^{f_b}$ contribution corresponding to 
$\alpha_r=0$.  
Also, because the EW calculation was performed with two-cutoff phase-space-slicing 
this EW contribution is finite.  Therefore, care is taken 
to only return values of $R_{\textnormal{EW}}^{f_b}$ which pass certain criteria, namely 
that they be away from the soft or collinear regions of the photon phase space. 
One can also consider subsets of the real corrections by adjusting the flag \texttt{QED} when \texttt{qnonr=0}.
The real EW contributions, up to the PDF factors, are proportional to subsets of the QED radiation 
matrix element squared as follows
\begin{align}
R_{\textnormal{EW}}^{f_b} \sim \sum_{f_b}\overline{\sum_{\text{\tiny{sp,col}}}}|M_{2\rightarrow3}|^2 &=
\sum_{f_b}\overline{\sum_{\text{\tiny{sp,col}}}}|M_{\text{IS}}+M_{\text{FS}}|^2_{f_b}\\
&= \sum_{f_b}\overline{\sum_{\text{\tiny{sp,col}}}}\underbrace{\left[
\underbrace{|M_{\text{IS}}|^2_{f_b}}_{\texttt{QED=1}} + 
\underbrace{|M_{\text{FS}}|^2_{f_b}}_{\texttt{QED=2}} + 
\underbrace{2\mathcal{R}e\left[M_{\text{IS}}M^*_{\text{FS}}\right]_{f_b}}_{\texttt{QED=3}} \right]}
_{\texttt{QED=4}} 
\end{align}
The integration of $R_{\textnormal{EW}}^{f_b}$ over the photon phase space is finite after applying soft and collinear cuts
as shown in Eq.~\ref{eq:main}. The dependence on the PSS parameters is canceled numerically between the contributions describing
soft, collinear and real hard photon radiation as discussed in Section~\ref{sec:checks}. The default values set in \texttt{/POWHEG-BOX/W\_EW-BW/virtual\_EW.f}
represent an optimal choice and should only be changed with care.

\end{document}